\documentclass[a4paper,10pt]{article}
\usepackage[utf8]{inputenc}
\usepackage{amsthm}
\usepackage{amsfonts}
\usepackage{amssymb}	
\usepackage{amsmath}
\allowdisplaybreaks
\usepackage{mathtools}
\usepackage[english]{babel}
\usepackage{color}
\usepackage{slashed}
\usepackage{enumerate}
\usepackage{graphicx}
\usepackage{systeme}
\usepackage{dsfont}
\usepackage{relsize}
\usepackage[margin=0.90in]{geometry}
\usepackage{float}
\usepackage{tikz}
\usepackage[labelformat=simple]{subcaption}

\usepackage{graphicx}
\usepackage{bmpsize}
\usepackage{epstopdf}
\usepackage{bbm}
\usepackage{xcolor,etoolbox}
\usepackage{titling}
\usepackage{authblk}
\newcommand*\samethanks[1][\value{footnote}]{\footnotemark[#1]}
\usepackage{blindtext,graphicx}
\usepackage[absolute]{textpos}
\usepackage{physics}
\usepackage[hang,flushmargin]{footmisc}

\usepackage{xfrac}
\usepackage{nicefrac}
\usepackage{esvect}
\usepackage{cases}
\usepackage{empheq}
\usepackage{stmaryrd}
\usepackage{cancel}
\usepackage[linguistics]{forest}
\usepackage{url}
\usepackage[font=small]{caption}
\usepackage[giveninits=true,sortcites=true,date=year,maxbibnames=99,doi=false,isbn=false,url=false,eprint=false]{biblatex}
\renewbibmacro{in:}{}
\DeclareFieldFormat{pages}{#1}
\AtEveryBibitem{%
  \clearlist{language}%
}
\usepackage{csquotes}

\usepackage{setspace}
\theoremstyle{definition}

\makeatletter
\newcommand{\pushright}[1]{\ifmeasuring@#1\else\omit\hfill$\displaystyle#1$\fi\ignorespaces}
\newcommand{\pushleft}[1]{\ifmeasuring@#1\else\omit$\displaystyle#1$\hfill\fi\ignorespaces}
\makeatother
\title{\vspace{-2cm}Analytical solution of the uniaxial extension problem for the relaxed micromorphic continuum and other generalized continua (including full derivations)}
\author{
	Gianluca Rizzi\thanks{corresponding author, GEOMAS, INSA-Lyon, Universit\'e de Lyon, 20 avenue Albert Einstein,	69621, \\ \hspace*{0.55cm} Villeurbanne cedex, France, gianluca.rizzi@insa-lyon.fr},
	\quad\quad
	Hassam Khan\thanks{Fakultät für Mathematik, Universität Duisburg-Essen, Thea-Leymann-Straße 9, 45127 Essen, Germany},
	\quad\quad
	Ionel-Dumitrel Ghiba\thanks{Department of Mathematics, Alexandru Ioan Cuza University of Ia\c si, Blvd. Carol I, no. 11, 700506 Ia\c si, Romania; and Octav Mayer Institute of Mathematics of the Romanian Academy, Ia\c si Branch, 700505 Ia\c si},
	\\
	Angela Madeo\samethanks[1]
	\quad and \quad
	Patrizio Neff\thanks{Head of Chair for Nonlinear Analysis and Modelling, Fakultät für Mathematik, Universität Duisburg-Essen, \\ \hspace*{0.55cm} Thea-Leymann-Straße 9, 45127 Essen, Germany}}

\thanksmarkseries{arabic}
\date{\today}
\begin{document}
\maketitle
\begin{abstract}
We derive analytical solutions for the uniaxial extension problem for the relaxed micromorphic continuum and other generalized continua.
These solutions may help in the identification of material parameters of generalized continua which are able to disclose size-effects.
\end{abstract}

\section{Introduction }
\label{sec:intro}
\begin{figure}[H]
	\centering
	\begin{subfigure}{0.49\textwidth}
	    \centering
		\includegraphics[width=\textwidth]{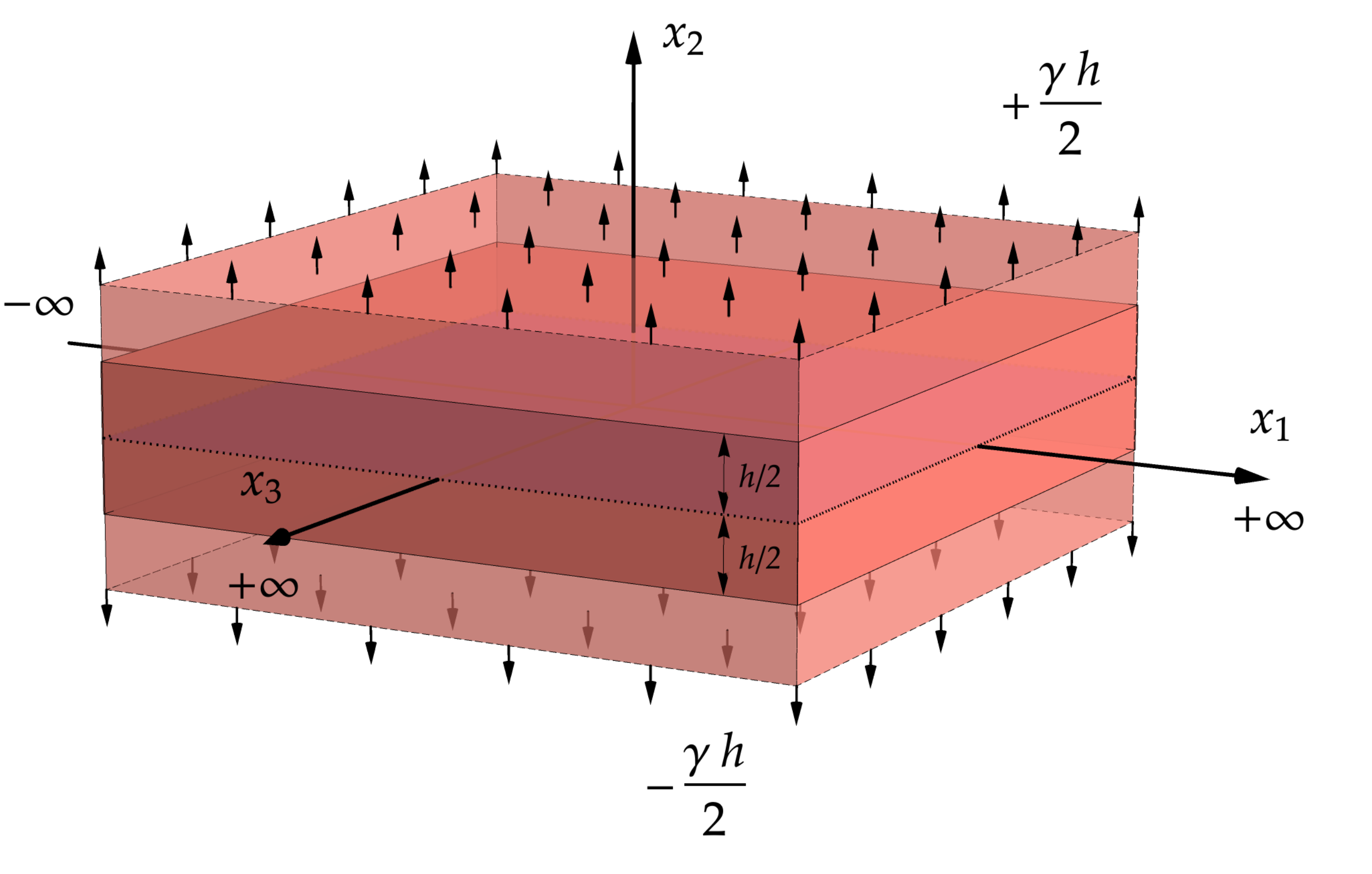}
	\end{subfigure}
	\hfill
	\begin{subfigure}{0.49\textwidth}
		\centering
		\includegraphics[width=\textwidth]{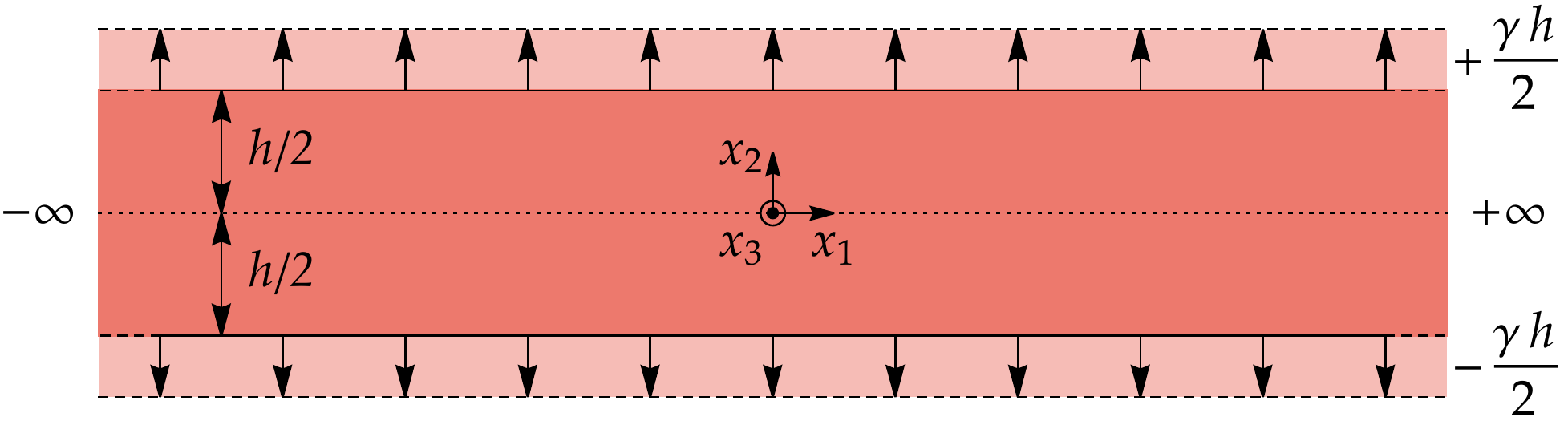}
	\end{subfigure}
	\caption{Sketch of an infinite stripe of thickness $h$ subjected to uniaxial extension boundary conditions.}
	\label{fig:intro}
\end{figure}
In this paper we continue our investigation of analytical solutions for isotropic the relaxed micromorphic model (and other isotropic generalized continuum models).
It follows our recent exposition of analytical solutions for the simple shear \cite{rizzi2021shear}, bending \cite{rizzi2021bending}, and torsion problem \cite{rizzi2021torsion,izadi2021torsional}.
Here, we consider the uniaxial extension problem, which, in classical isotropic linear elasticity, allows to determine the size-independent longitudinal modulus $M_{\text{macro}}=\lambda_{\text{macro}}+2\mu_{\text{macro}}$.

Here, we show the genealogy tree of the generalized continuum models:
\begin{center}
\begin{forest}
    [
    {
    \textit{classical micromorphic}
    \\
    $
    \displaystyle \min_{\boldsymbol{u},\boldsymbol{P}}
    \Big[
    W \left(\boldsymbol{\text{D}u},\boldsymbol{P}, \boldsymbol{\text{D}P} \right)
    \Big]
    $
    }
        [
        {
        \textit{micro-strain} ($\boldsymbol{P} = \boldsymbol{S}$)
        \\
        $
        \displaystyle \min_{\boldsymbol{u},\boldsymbol{S}}
        \Big[
        W \left(\boldsymbol{\text{D}u},\boldsymbol{S}, \boldsymbol{\text{D}S} \right)
        \Big]
        $
        }
            [
            {
            \textit{strain gradient} ($\boldsymbol{S} = \text{sym}~\boldsymbol{\text{D}u}$)
            \\
            $
            \displaystyle \min_{\boldsymbol{u}}
            \Big[
            W \left(\boldsymbol{\text{D}u}, \boldsymbol{\text{D} \left(\text{sym}~\boldsymbol{\text{D}u}\right)} \right)
            \Big]
            $
            }
            ]
        ]
        [
        {
        \qquad
        \textit{relaxed micromorphic}
        \qquad
        \\
        \qquad
        $
        \displaystyle \min_{\boldsymbol{u},\boldsymbol{P}}
        \Big[
        W \left(\boldsymbol{\text{D}u},\boldsymbol{P}, \boldsymbol{\text{Curl}P} \right)
        \Big]
        $
        \qquad
        }
            [
            {
            \qquad\qquad
            \textit{micro-stretch} ($\boldsymbol{P} = \boldsymbol{A} + \omega \boldsymbol{\mathbbm{1}}$)
            \qquad\qquad
            \\
            $
            \displaystyle \min_{\boldsymbol{u},\boldsymbol{A},\omega}
            \Big[
            W \left(\boldsymbol{\text{D}u},\boldsymbol{A} , \omega \boldsymbol{\mathbbm{1}}, \boldsymbol{\text{Curl} \left( \boldsymbol{A} + \omega \boldsymbol{\mathbbm{1}} \right)} \right)
            \Big]
            $
            }
                [
                {
                \qquad\qquad\qquad
                \textit{Cosserat} ($\boldsymbol{P} = \boldsymbol{A}$)
                \qquad\qquad\qquad
                \\
                $
                \displaystyle \min_{\boldsymbol{u},\boldsymbol{A}}
                \Big[
                W \left(\boldsymbol{\text{D}u},\boldsymbol{A}, \boldsymbol{\text{Curl} \, \boldsymbol{A} } \right)
                \Big]
                $
                }
                    [
                    {
                    \textit{couple stress} ($\boldsymbol{A} = \text{skew}~\boldsymbol{\text{D}u}$)
                    \\
                    $
                    \displaystyle \min_{\boldsymbol{u}}
                    \Big[
                    W \left(\boldsymbol{\text{D}u}, \boldsymbol{\text{Curl} \left(\text{skew}~\boldsymbol{\text{D}u}\right)} \right)
                    \Big]
                    $
                    }
                        [
                        {
                        \textit{skew symmetric couple stress}
                        \\
                        $
                        \displaystyle \min_{\boldsymbol{u}}
                        \Big[
                        W \left(\boldsymbol{\text{D}u}, \text{skew}~\boldsymbol{\text{Curl} \left(\text{skew}~\boldsymbol{\text{D}u}\right)} \right)
                        \Big]
                        $
                        }
                        ]
                        [
                        {
                        \textit{modified couple stress}
                        \\
                        $
                        \displaystyle \min_{\boldsymbol{u}}
                        \Big[
                        W \left(\boldsymbol{\text{D}u}, \text{sym}~\boldsymbol{\text{Curl} \left(\text{skew}~\boldsymbol{\text{D}u}\right)} \right)
                        \Big]
                        $
                        }
                        ]
                    ]
                ]
                [
                {
                \qquad\qquad\qquad
                \textit{micro-void} ($\boldsymbol{P} = \omega \boldsymbol{\mathbbm{1}}$)
                \qquad\qquad\qquad
                \\
                $
                \displaystyle \min_{\boldsymbol{u},\omega}
                \Big[
                W \left(\boldsymbol{\text{D}u},\omega, \boldsymbol{\text{Curl} \left( \omega \boldsymbol{\mathbbm{1}} \right)} \right)
                \Big]
                $
                }
                [
                {
                $
                \boxed{
                \begin{array}{rl}
                    \boldsymbol{u}&: \Omega \subset \mathbb{R}^{3} \to \mathbb{R}^{3} \, ,
                    \\*
                    \boldsymbol{P}&: \Omega \subset \mathbb{R}^{3} \to \mathbb{R}^{3\times 3} \, ,
                    \\*
                    \boldsymbol{A}&: \Omega \subset \mathbb{R}^{3} \to \mathfrak{so}(3) \, ,
                    \\*
                    \boldsymbol{S}&: \Omega \subset \mathbb{R}^{3} \to \text{Sym}(3) \, ,
                    \\*
                    \omega&: \Omega \subset \mathbb{R}^{3} \to \mathbb{R} \, ,
                \end{array}
                }
                $
                },no edge
                ]
                ]
            ]
        ]
        [
        {
        \textit{second gradient} ($\boldsymbol{P} = \boldsymbol{\text{D}u}$)
        \\
        $
        \displaystyle \min_{\boldsymbol{u}}
        \Big[
        W \left(\boldsymbol{\text{D}u}, \boldsymbol{\text{D}^2 u} \right)
        \Big]
        $
        }
        ]
    ]
\end{forest}
\end{center}

The strain gradient theory and second gradient theory are equivalent \cite{mindlin1964micro,altenbach2019higher}, and contain additionally the couple stress theory as a special case.
Using the ${\rm Curl}$ as primary differential operator for the curvature terms allows a neat unification of concepts.

For some of the traditional models, uniaxial extension gives still rise to size-effects in the sense that thinner samples are comparatively stiffer.
In that case, the inhomogeneous response is triggered by the boundary conditions for the additional kinematic fields which are applied at the upper and lower surface.
We refer the reader to the introduction of \cite{rizzi2021shear,rizzi2021bending,rizzi2021torsion,shaat2020review} concerning the relevance of the scientific question as well as its importance for the determination of material parameters for generalized continua \cite{shekarchizadehinverse}.
Indeed, the obtained analytical formulas can be used to determine size-dependent and size-independent material parameters.
The notation follows that of \cite{rizzi2021shear,rizzi2021bending,rizzi2021torsion}.
We recapitulate shortly.

The paper is now structured as follows.
We start with a recapitulation of the uniaxial extension problem in the classical linear elasticity.
The solution is homogeneous and uniquely determines the longitudinal modulus $M_{\text{macro}}=\lambda_{\text{macro}}+2\mu_{\text{macro}}$.
Then, we consider the isotropic relaxed micromorphic continuum.
The boundary conditions for the additional non-symmetric micro-distortion field $\boldsymbol{P}$ derives from the so-called consistent coupling conditions
\begin{align}
    \text{D}\boldsymbol{u}(\boldsymbol{x})\times \boldsymbol{\nu} = \boldsymbol{P}(\boldsymbol{x})\times \boldsymbol{\nu}
    \, ,
    \qquad\qquad\qquad
    \boldsymbol{x} \in \Gamma
    \, ,
\end{align}
where $\boldsymbol{\nu}$ is the normal unit vector to the upper and lower surface.
It turns out that for zero Poisson modulus on the micro- and meso-scale, $\nu_{\text{micro}}=\nu_{\text{e}}=0$, respectively, the solution remains homogeneous and no size-effects is observed.
In the case with arbitrary $\nu_{\text{micro}},\nu_{\text{e}}\in \left[ -1,1/2\right]$ the solution will be inhomogeneous and size-effects appear.
The limiting stiffness as the ratio between the thickness and the characteristic length tends to zero ($h/L_{\text{c}}\to0$) is given by $\overline{M} = \frac{M_{\text{e}} \, M_{\text{micro}}}{M_{\text{e}} + M_{\text{micro}}}$ which is both smaller then $M_{\text{micro}}=\lambda_{\text{micro}} + 2\mu_{\text{micro}}$ and $M_{\text{e}}$ as well greater than $M_{\text{macro}}=\lambda_{\text{macro}} + 2\mu_{\text{macro}}$.

\allowdisplaybreaks
\subsection{Notation}
\label{sec:notation}
We define the scalar product $\langle \boldsymbol{a},\boldsymbol{b} \rangle \coloneqq \sum_{i=1}^n a_i\,b_i \in \mathbb{R}$ for vectors $a,b\in\mathbb{R}^n$, the dyadic product  $\boldsymbol{a}\otimes \boldsymbol{b} \coloneqq \left(a_i\,b_j\right)_{i,j=1,\ldots,n}\in \mathbb{R}^{n\times n}$ and the euclidean norm  $\norm{\boldsymbol{a}}^2\coloneqq\langle \boldsymbol{a},\boldsymbol{a} \rangle$.
We define the scalar product $\langle \boldsymbol{P},\boldsymbol{Q} \rangle \coloneqq\sum_{i,j=1}^n P_{ij}\,Q_{ij} \in \mathbb{R}$ and the Frobenius-norm $\norm{\boldsymbol{P}}^2\coloneqq\langle \boldsymbol{P},\boldsymbol{P} \rangle$ for tensors $\boldsymbol{P},\boldsymbol{Q}\in\mathbb{R}^{n\times n}$ in the same way.
Moreover, $\boldsymbol{P}^T\coloneqq (P_{ji})_{i,j=1,\ldots,n}$ denotes the transposition of the matrix $\boldsymbol{P}=(P_{ij})_{i,j=1,\ldots,n}$, which decomposes orthogonally into the skew-symmetric part $\text{skew} \, \boldsymbol{P} \coloneqq \frac{1}{2} (\boldsymbol{P}-\boldsymbol{P}^T )$ and the symmetric part $\text{sym} \, \boldsymbol{P} \coloneqq \frac{1}{2} (\boldsymbol{P}+\boldsymbol{P}^T)$.
The identity matrix is denoted by $\boldsymbol{\mathbbm{1}}$, so that the trace of a matrix $\boldsymbol{P}$ is given by \ $\tr \boldsymbol{P} \coloneqq \langle \boldsymbol{P},\boldsymbol{\mathbbm{1}} \rangle$, while the deviatoric component of a matrix is given by $\text{dev} \, \boldsymbol{P} \coloneqq \boldsymbol{P} - \frac{\tr \left( \boldsymbol{P}\right)}{3} \, \boldsymbol{\mathbbm{1}}$.
Given this, the  orthogonal decomposition possible for a matrix is $\boldsymbol{P} = \text{dev} \,\text{sym} \, \boldsymbol{P} + \text{skew} \, \boldsymbol{P} + \frac{\tr \left( \boldsymbol{P}\right)}{3} \, \boldsymbol{\mathbbm{1}}$.
The Lie-Algebra of skew-symmetric matrices is denoted by $\mathfrak{so}(3)\coloneqq \{\boldsymbol{A}\in\mathbb{R}^{3\times 3}\mid \boldsymbol{A}^T = -\boldsymbol{A}\}$,
while the vector space of symmetric matrices  $\text{Sym}(3)\coloneqq \{\boldsymbol{S}\in\mathbb{R}^{3\times 3}\mid \boldsymbol{S}^T = \boldsymbol{S}\}$.
The Jacobian matrix D$\boldsymbol{u}$ and the curl for a vector field $\boldsymbol{u}$ are defined as
\begin{equation}
\boldsymbol{\text{D}u}
=\!
\left(
\begin{array}{ccc}
u_{1,1} & u_{1,2} & u_{1,3} \\
u_{2,1} & u_{2,2} & u_{2,3} \\
u_{3,1} & u_{3,2} & u_{3,3}
\end{array}
\right)\, ,
\qquad
\text{curl} \, \boldsymbol{u} = \boldsymbol{\nabla} \times \boldsymbol{u}
=
\left(
\begin{array}{ccc}
u_{3,2} - u_{2,3}  \\
u_{1,3} - u_{3,1}  \\
u_{2,1} - u_{1,2}
\end{array}
\right)
\, .
\end{equation}
where $\times$ denotes the cross product in $\mathbb{R}^3$.
We also introduce the $\text{Curl}$ and the $\text{Div}$ operators of the $3\times 3$ matrix field $\boldsymbol{P}$ as
\begin{equation}
\text{Curl} \, \boldsymbol{P}
=\!
\left(
\begin{array}{c}
(\text{curl}\left( P_{11} , \right. P_{12} , \left. P_{13} \right)^{{T}})^T \\
(\text{curl}\left( P_{21} , \right. P_{22} , \left. P_{23} \right)^{T})^T \\
(\text{curl}\left( P_{31} , \right. P_{32} , \left. P_{33} \right)^{T})^T
\end{array}
\right) \!,
\qquad
\text{Div}  \, \boldsymbol{P}
=\!
\left(
\begin{array}{c}
\text{div}\left( P_{11} , \right. P_{12} , \left. P_{13} \right)^{T} \\
\text{div}\left( P_{21} , \right. P_{22} , \left. P_{23} \right)^{T} \\
\text{div}\left( P_{31} , \right. P_{32} , \left. P_{33} \right)^{T}
\end{array}
\right) \, .
\end{equation}
The cross product between a second order tensor and a vector is also needed and is defined row-wise as follow
\begin{equation}
\boldsymbol{m} \times \boldsymbol{b} =
\left(
\begin{array}{ccc}
(b \times (m_{11},m_{12},m_{13})^{T})^T \\
(b \times (m_{21},m_{22},m_{23})^{T})^T \\
(b \times (m_{31},m_{32},m_{33})^{T})^T \\
\end{array}
\right) =
\boldsymbol{m} \cdot \boldsymbol{\epsilon} \cdot \boldsymbol{b} =
m_{ik} \, \epsilon_{kjh} \, b_{h} 
\, ,
\end{equation}
where $\boldsymbol{m} \in \mathbb{R}^{3\times 3}$, $\boldsymbol{b} \in \mathbb{R}^{3}$, and $\boldsymbol{\epsilon}$ is the Levi-Civita tensor.
Using the one-to-one map $\text{axl}:\mathfrak{so}(3)\to\mathbb{R}^3$ we have
\begin{align}
\boldsymbol{A} \, \boldsymbol{b} =\text{axl}(\boldsymbol{A})\times \boldsymbol{b} \quad \forall\, \boldsymbol{A}\in\mathfrak{so}(3) \, ,
\quad
\boldsymbol{b}\in\mathbb{R}^3.
\label{eq:Aanti_axl}
\end{align}
The inverse of  axl is denoted by Anti: $\mathbb{R}^3\to \mathfrak{so}(3)$.
\section{Uniaxial extension problem for the isotropic Cauchy continuum}
\label{sec:Cau}
The strain energy density for an isotropic Cauchy continuum is
\begin{equation}
W \left(\text{D}\boldsymbol{u}\right) = 
\mu_{\text{macro}} \left\lVert \text{sym} \text{D}\boldsymbol{u} \right\rVert^{2} + 
\dfrac{\lambda_{\text{macro}}}{2} \text{tr}^2\left(\text{D}\boldsymbol{u}\right)
\, ,
\label{eq:energyCau}
\end{equation}
while the equilibrium equations without body forces are
\begin{equation}
\text{Div}\left[ 
2\,\mu_{\text{macro}}\,\text{sym} \text{D}\boldsymbol{u}
+ \lambda_{\text{macro}}\,\text{tr}\left(\text{D}\boldsymbol{u}\right)  \boldsymbol{\mathbbm{1}} 
\right] 
= \boldsymbol{0}.
\label{eq:equiCau}
\end{equation}
Since the uniaxial extensional problem is symmetric with respect to the $x_2$-axis, there will be no dependence of the solution on $x_1$ and $x_3$.
The boundary conditions for the uniaxial extension problem are (see Fig.~\ref{fig:intro})
\begin{align}
    u_{2}(x_2=\pm h/2) = \pm \frac{\boldsymbol{\gamma} \, h}{2}
    \, .
\end{align}
The homogeneous displacement field solution $u_{2} (x_2)$, the gradient of the displacement $\text{D}\boldsymbol{u}(x_2)$, and the strain energy $W(\boldsymbol{\gamma})$ for the uniaxial extension problem are
\begin{align}
u_{2} (x_2) &=
\boldsymbol{\gamma} \, x_{2} \, ,
\qquad\qquad\qquad
\text{D}\boldsymbol{u}(x_2) =
	\left(
	\begin{array}{ccc}
		0 & 0 & 0 \\
		0 & \boldsymbol{\gamma} & 0 \\
		0 & 0 & 0  \\
	\end{array}
	\right)
\, ,
\label{eq:class_solu}
\\*
W(\boldsymbol{\gamma})
&=
\int_{-h/2}^{h/2} W(\text{D}\boldsymbol{u})
=
\frac{1}{2}
\left(\lambda_{\text{macro}}+2\mu_{\text{macro}}\right) h \, \boldsymbol{\gamma}^2
=
\frac{1}{2} \, M_{\text{macro}} \, h \, \boldsymbol{\gamma}^2
\, ,
\notag
\end{align}
where
\begin{align}
M_{\text{macro}} = \lambda_{\text{macro}}+2\mu_{\text{macro}}    
\end{align}
is the extensional stiffness (or pressure-wave modulus, longitudinal modulus).

Here and in the remainder of this work, the elastic coefficients $\mu_i,\lambda_i$ are expressed in [MPa], the coefficients $a_i$ and the intensity of the displacement $\boldsymbol{\gamma}$ are dimensionless, the characteristic lengths $L_{\text{c}}$ and the height $h$ are expressed in meter [m].
\section{Uniaxial extension problem for the isotropic relaxed micromorphic model}
\label{sec:Relaxed_micro}

The  general expression of the strain energy for the isotropic relaxed micromorphic continuum is
\begin{align}
	W \left(\boldsymbol{\text{D}u}, \boldsymbol{P},\text{Curl}\,\boldsymbol{P}\right)
	=
	&
	\, \mu_{\text{e}} \left\lVert \text{sym} \left(\boldsymbol{\text{D}u} - \boldsymbol{P} \right) \right\rVert^{2}
	+ \frac{\lambda_{\text{e}}}{2} \text{tr}^2 \left(\boldsymbol{\text{D}u} - \boldsymbol{P} \right) 
	+ \mu_{\text{c}} \left\lVert \text{skew} \left(\boldsymbol{\text{D}u} - \boldsymbol{P} \right) \right\rVert^{2}
	\notag
	\\*
	&
	+ \mu_{\text{micro}} \left\lVert \text{sym}\,\boldsymbol{P} \right\rVert^{2}
	+ \frac{\lambda_{\text{micro}}}{2} \text{tr}^2 \left(\boldsymbol{P} \right)
	\label{eq:energy_RM}
	\\*
	&
	+ \frac{\mu \,L_{\text{c}}^2 }{2} \,
	\left(
	a_1 \, \left\lVert \text{dev sym} \, \text{Curl} \, \boldsymbol{P}\right\rVert^2 +
	a_2 \, \left\lVert \text{skew} \,  \text{Curl} \, \boldsymbol{P}\right\rVert^2 +
	\frac{a_3}{3} \, \text{tr}^2 \left(\text{Curl} \, \boldsymbol{P}\right)
	\right)
	\notag
	\, ,
	\notag
\end{align}
and the strictly positive definiteness conditions are
\!\!\!
\footnote{
Note that the model has a unique solution including the case of a Cosserat couple modulus $\mu_{\text{c}}=0$.
}
\begin{gather}
\label{eq:posi_defe_RM}
    \mu_{\text{e}} > 0,
    \qquad
    \kappa_{\text{e}} = \lambda_{\text{e}}+2/3 \, \mu_{\text{e}} > 0,
    \qquad
    \mu_{\text{micro}} > 0,
    \qquad
    \kappa_{\text{micro}}=\lambda_{\text{micro}}+2/3 \, \mu_{\text{micro}} > 0,
    \\*
    \mu_{\text{c}} > 0,
    \qquad
    \mu > 0,
    \qquad
    L_{\text{c}} > 0,
    \qquad (a_1,a_2,a_3) > 0
    \, .
    \notag
\end{gather}
where we have the parameters related to the meso-scale, the parameters related to the micro-scale, the Cosserat couple modulus, the proportionality stiffness parameter, the characteristic length, and the three dimensionless general isotropic curvature parameters, respectively.
This energy expression represents the  most general isotropic form possible for the relaxed micromorphic model.
In the absence of body forces, the equilibrium equations  are then
\begin{align}
	\text{Div}\overbrace{\left[2\mu_{\text{e}}\,\text{sym} \left(\boldsymbol{\text{D}u} - \boldsymbol{P} \right) + \lambda_{\text{e}} \text{tr} \left(\boldsymbol{\text{D}u} - \boldsymbol{P} \right) \boldsymbol{\mathbbm{1}}
		+ 2\mu_{\text{c}}\,\text{skew} \left(\boldsymbol{\text{D}u} - \boldsymbol{P} \right)\right]}^{\mathlarger{\widetilde{\sigma}}\coloneqq}
	&= \boldsymbol{0},
	\notag
	\\*
	\widetilde{\sigma}
	- 2 \mu_{\text{micro}}\,\text{sym}\,\boldsymbol{P}
	- \lambda_{\text{micro}} \text{tr} \left(\boldsymbol{P}\right) \boldsymbol{\mathbbm{1}}
	\hspace{8cm}
	\label{eq:equi_RM}
	\\*
	- \mu \, L_{\text{c}}^{2} \, \text{Curl}
	\big(
	\underbrace{a_1 \, \text{dev sym} \, \text{Curl} \, \boldsymbol{P} +
	a_2 \, \text{skew} \, \text{Curl} \, \boldsymbol{P} +
	a_3 \, \text{tr} \left(\text{Curl} \, \boldsymbol{P}\right)}_{
	\boldsymbol{m}}
	\big)
	&= \boldsymbol{0} \, .
	\notag
\end{align}
The ansatz for the micro-distortion $\boldsymbol{P}(x_2)$, the displacement $\boldsymbol{u}(x_2)$, and consequently the gradient of the displacement $\text{D}\boldsymbol{u}(x_2)$ is
\begin{align}
	\boldsymbol{u}(x_2) &=
	\left(
	\begin{array}{c}
		0 \\
		u_{2}(x_{2}) \\
		0 
	\end{array}
	\right)
	\, ,
	\label{eq:ansatz_RM}
	\qquad\qquad
	\boldsymbol{P}(x_2) =
	\left(
	\begin{array}{ccc}
		P_{11}(x_2) & 0 & 0 \\
		0 & P_{22}(x_2) & 0 \\
		0 & 0 & P_{33}(x_2)  \\
	\end{array}
	\right)
	\, ,
	\\*
	\text{D}\boldsymbol{u}(x_2) &=
	\left(
	\begin{array}{ccc}
		0 & 0 & 0 \\
		0 & u_{2,2}(x_2) & 0 \\
		0 & 0 & 0  \\
	\end{array}
	\right)
	\, .
	\notag
\end{align}
It is important to underline that, given the subsequent ansatz (\ref{eq:ansatz_RM}), it holds that $\text{tr} \left(\text{Curl} \, \boldsymbol{P}\right)=0$. This reduces immediately the number of curvature parameters appearing in the uniaxial extension solution.

The boundary conditions for the uniaxial extension are
\begin{align}
u_{2}(x_{2} = \pm h/2) = \pm \frac{\boldsymbol{\gamma} \, h}{2}
\, , 
\qquad\qquad
P_{11}(x_{2} = \pm h/2) = 0
\, ,
\qquad\qquad
P_{33}(x_{2} = \pm h/2)  = 0
\, .
\label{eq:BC_RM}
\end{align}
Here, the constraint on the components of $\boldsymbol{P}$ is given by the \textit{consistent coupling boundary condition}
\begin{align}
\boldsymbol{P}\times \boldsymbol{\nu}
=
\text{D}\boldsymbol{u}\times \boldsymbol{\nu}
\, ,
\qquad\qquad\qquad
\left(
\begin{array}{ccc}
0 & 0 & 0 \\
0 & 0 & 0 \\
0 & 0 & 0 \\
\end{array}
\right)
=
\left(
\begin{array}{ccc}
 0 & 0 & P_{11} \\
 0 & 0 & 0 \\
 -P_{33} & 0 & 0 \\
\end{array}
\right)
\, ,
\end{align}
where $\boldsymbol{\nu}$ is the normal unit vector to the upper and lower surface.

After substituting the ansatz (\ref{eq:ansatz_RM}) into the equilibrium equation (\ref{eq:equi_RM}) we obtain the following four differential equations
\begin{align}
    M_{\text{e}} \left(u_{2}''(x_{2})-P_{22}'(x_{2})\right)
    -\lambda_{\text{e}} \left(P_{11}'(x_{2})+P_{33}'(x_{2})\right)
    & =
    0 \, ,
    \notag
    \\*[2mm]
    \frac{1}{2} \mu \, L_{\text{c}}^2 \left((a_{1}+a_{2}) P_{11}''(x_{2})+(a_{2}-a_{1}) P_{33}''(x_{2})\right)
    \notag
    \\*
    -(M_{\text{e}}+M_{\text{micro}}) P_{11}(x_{2})
    -(\lambda_{\text{e}}+\lambda_{\text{micro}}) (P_{22}(x_{2})+P_{33}(x_{2}))
    +\lambda_{\text{e}} u_{2}'(x_{2})
    & =
    0 \, ,
    \label{eq:equi_RM_2}
    \\*[2mm]
    -(M_{\text{e}}+M_{\text{micro}}) P_{22}(x_{2})+M_{\text{e}} u_{2}'(x_{2})
    -(\lambda_{\text{e}}+\lambda_{\text{micro}}) (P_{11}(x_{2})+P_{33}(x_{2}))
    & =
    0 \, ,
    \notag
    \\*[2mm]
    \frac{1}{2} \mu \, L_{\text{c}}^2 \left((a_{2}-a_{1}) P_{11}''(x_{2})+(a_{1}+a_{2}) P_{33}''(x_{2})\right)
    \notag
    \\*
    -(M_{\text{e}}+M_{\text{micro}}) P_{33}(x_{2})
    -(\lambda_{\text{e}}+\lambda_{\text{micro}}) (P_{11}(x_{2})+P_{22}(x_{2}))+\lambda_{\text{e}} u_{2}'(x_{2})
    & =
    0\, ,
    \notag
\end{align}
where $M_{\text{e}}=\lambda_{\text{e}}+2\mu_{\text{e}}$ and $M_{\text{micro}}=\lambda_{\text{micro}}+2\mu_{\text{micro}}$.
Being careful of substituting the system of differential equation with one in which eq.(\ref{eq:equi_RM_2})$_2$ and eq.(\ref{eq:equi_RM_2})$_4$ are replaced with their sum and their difference, respectively, we have
\begin{align}
    M_{\text{e}} \left(u_{2}''(x_{2})-P_{22}'(x_{2})\right)-\lambda_{\text{e}} f_{p}'(x_{2}) 
    =
    0 \, ,
    \notag
    \\*[2mm]
    a_{2} \, \mu \, L_{\text{c}}^2 \, f_{p}''(x_{2})
    - (M_{\text{e}}+\lambda_{\text{e}}+M_{\text{micro}}+\lambda_{\text{micro}}) f_{p}(x_{2})
    - 2 (\lambda_{\text{e}}+\lambda_{\text{micro}}) P_{22}(x_{2})
    + 2 \lambda_{\text{e}} u_{2}'(x_{2}) 
    =
    0 \, ,
    \label{eq:equi_RM_3}
    \\*[2mm]
    - (M_{\text{e}}+M_{\text{micro}}) P_{22}(x_{2})
    + M_{\text{e}} \, u_{2}'(x_{2})
    - (\lambda_{\text{e}}+\lambda_{\text{micro}}) f_{p}(x_{2})
     =
     0 \, ,
    \notag
    \\*[2mm]
    a_{1} \, \mu \, L_{\text{c}}^2 \, f_{m}''(x_{2})
    -
    (M_{\text{micro}} + M_{\text{e}} - \lambda_{\text{e}} - \lambda_{\text{micro}})f_{m}(x_{2})
    =
    0 \, ,
    \notag
\end{align}
where $f_{p} (x_2)\coloneqq P_{11}(x_2)+P_{33}(x_2)$ and $f_{m} (x_2)\coloneqq P_{11}(x_2)-P_{33}(x_2)$.
It is highlighted that eq.(\ref{eq:equi_RM_3})$_4$ is a homogeneous second order differential equation depending only on $f_{m}(x_2)$ with homogeneous boundary conditions eq.(\ref{eq:BC_RM}).

The fact that eq.(\ref{eq:equi_RM_3})$_4$ is an independent equation has its meaning in the symmetry constraint of the uniaxial extensional problem in the direction along the $x_2$- and $x_3$-axis, which requires that ${P_{11}(x_2)=P_{33}(x_2)}$.
From eq.(\ref{eq:equi_RM_3}) it is possible to obtain the following relation between $P_{22}(x_2)$ and $u_2(x_2)$
\begin{align}
    P_{22}(x_{2})
    =
    \frac{
    M_{\text{e}} \,  u_{2}'(x_{2})
    - (\lambda_{\text{e}}+\lambda_{\text{micro}}) f_{p}(x_{2})
    }{
    M_{\text{e}}+M_{\text{micro}}
    }
    \, ,
    \label{eq:rel_P11_du}
\end{align}
which, after substituting it back into eq.(\ref{eq:equi_RM_3}), allows us to obtain the following system of three second order differential equations in $u_2(x_2)$, $P_{22}(x_2)$, and $f_{p}(x_2)$
\begin{align}
    z_{1} \, f_{p}'(x_{2})
    + z_{2} \, u_{2}''(x_{2}) 
    =
    0\, ,
    \notag
    \\*
    a_{2} \, \mu \, L_{\text{c}}^2 \, f_{p}''(x_{2})
    - z_{3} \, f_{p}(x_{2})
    - 2 z_{1} \, u_{2}'(x_{2})
    =
    0\, ,
    \label{eq:equi_RM_4}
    \\*
    a_{1} \, \mu \, L_{\text{c}}^2 \, f_{m}''(x_{2})
    - (M_{\text{e}} + M_{\text{micro}} - \lambda_{\text{e}} - \lambda_{\text{micro}}) f_{m}(x_{2})
    =
    0\, ,
    \notag
    \end{align}
where
    \begin{align}
    z_1\coloneqq
    &
    \frac{
    M_{\text{e}} \lambda_{\text{micro}}-\lambda_{\text{e}} M_{\text{micro}}
    }{
    M_{\text{e}}+M_{\text{micro}}
    }
    \, ,
    \qquad\qquad\qquad
    z_2\coloneqq
    \frac{
    M_{\text{e}} M_{\text{micro}}
    }{
    M_{\text{e}}+M_{\text{micro}}
    }
    \, ,
    \label{eq:coeff_z_RM}
    \\*
    z_3\coloneqq
    &
    \frac{
    \left(M_{\text{e}}-\lambda_{\text{e}}+M_{\text{micro}}-\lambda_{\text{micro}}\right)
    \left(M_{\text{e}}+2 \lambda_{\text{e}}+M_{\text{micro}} +2 \lambda_{\text{micro}}\right)
    }{
    M_{\text{e}}+M_{\text{micro}}
    }
    \, .
    \notag
\end{align}
It is highlighted that due to the positive definiteness conditions (\ref{eq:posi_defe_RM}), $(z_2,z_3)>0$ and $z_1=0$ if and only if $\lambda_{\text{micro}}=\lambda_{\text{e}}=0$ (zero Poisson's ratio case which is studied in Sec.~\ref{sec:Relaxed_micro_2}) and $\frac{M_{\text{micro}}}{M_{\text{e}}}=\frac{\lambda_{\text{micro}}}{\lambda_{\text{e}}}$.
If $z_1$ is zero eqs.(\ref{eq:equi_RM_4}) uncouples completely into three independent differential equations in $u_2$, $f_{\text{p}}$, and $f_{\text{m}}$ respectively.

After applying the boundary conditions eqs.(\ref{eq:BC_RM}), the solution in terms of $u_2(x_2)$, $P_{11}(x_2)$, $P_{22}(x_2)$, and $P_{33}(x_2)$ of the system eqs.(\ref{eq:equi_RM_4}) is
\!\!\!
\footnote{$\sech(x) = 1/cosh(x)$.}
\begin{align}
    u_{2} (x_{2})
    &=
    \frac{
    \frac{2 x_{2}}{h}
    -\frac{4 z_{1}^2}{f_{1} z_{2} z_{3}}
    \, \sech\left(\frac{f_{1} h}{2 L_{\text{c}}}\right) \sinh \left(\frac{f_{1} x_{2}}{L_{\text{c}}}\right)
    \frac{L_{\text{c}}}{h}
    }{
    1-\frac{4 z_{1}^2}{f_{1} z_{2} z_{3}}
    \tanh \left(\frac{f_{1} h}{2 L_{\text{c}}}\right)
    \frac{L_{\text{c}}}{h}
    }
    \frac{\boldsymbol{\gamma}  h}{2}
    \, ,
    \notag
    \\*
    P_{22} (x_{2})
    &=
    \frac{
    M_{\text{e}}
    +
    2 \frac{z_1}{z_3}
    \left( \lambda_{\text{e}} + \lambda_{\text{micro}} \right)
    -
    \frac{z_1}{z_3}
    \left(
    M_{\text{e}}\frac{2z_1}{z_2}
    +2 \left( \lambda_{\text{e}} + \lambda_{\text{micro}} \right)
    \cosh\left(\frac{f_{1} h}{L_{\text{c}}}\right)
    \sech\left(\frac{f_{1} h}{2 L_{\text{c}}}\right)
    \right)
    }{
    \left(
    M_{\text{e}} + M_{\text{micro}}
    \right)
    \left(
    1-\frac{4 z_{1}^2}{f_{1} z_{2} z_{3}}
    \tanh \left(\frac{f_{1} h}{2 L_{\text{c}}}\right)
    \frac{L_{\text{c}}}{h}
    \right)
    }
    \,
    \boldsymbol{\gamma}
    \, ,
    \label{eq:solu_RM}
    \\*
    P_{11} (x_{2}) 
    &=
    P_{33} (x_{2})
    = 
    \frac{
    \frac{z_{1}}{z_{3}}
    \left(\sech\left(\frac{f_{1} h}{2 L_{\text{c}}}\right) \cosh \left(\frac{f_{1} x_{2}}{L_{\text{c}}}\right)-1\right)
    }{
    1-\frac{4 z_{1}^2}{f_{1} z_{2} z_{3}}
    \tanh \left(\frac{f_{1} h}{2 L_{\text{c}}}\right)
    \frac{L_{\text{c}}}{h}
    }
    \, 
    \boldsymbol{\gamma}
    \, ,
    \qquad\qquad
    f_{1}
    \coloneqq
    \sqrt{\frac{z_{2} \, z_{3} - 2 z_{1}^2}{ \mu \, a_{2} \, z_{2}}}
    \, .
    \notag
\end{align}
In the above expressions all the quantities are real and well defined due to the positive definiteness conditions eq.(\ref{eq:posi_defe_RM}). Indeed, since 
		 the coefficients $z_1$, $z_2$, and $z_3$ may be rewritten  in terms of the meso and micro bulk and shear modulus as
		\begin{gather}
		z_1 \coloneqq \frac{6 \kappa_{\text{micro}} \mu_{\text{e}}-6 \kappa_{\text{e}} \mu_{\text{micro}}}{3 \kappa_{\text{e}}+3 \kappa_{\text{micro}}+4 (\mu_{\text{e}}+\mu_{\text{micro}})} \, ,\notag
		\qquad
		z_2 \coloneqq \frac{(3 \kappa_{\text{e}}+4 \mu_{\text{e}}) (3 \kappa_{\text{micro}}+4 \mu_{\text{micro}})}{9 \kappa_{\text{e}}+9 \kappa_{\text{micro}}+12 (\mu_{\text{e}}+\mu_{\text{micro}})} \, ,
		\\*
		z_3 \coloneqq \frac{18 (\kappa_{\text{e}}+\kappa_{\text{micro}}) (\mu_{\text{e}}+\mu_{\text{micro}})}{3 \kappa_{\text{e}}+3 \kappa_{\text{micro}}+4 (\mu_{\text{e}}+\mu_{\text{micro}})} \, ,\notag
		\end{gather}
	 we can write the expression of $f_1$ as follows
		\begin{align}
		f_1 \coloneqq
		\sqrt{
			\frac{
				6 \kappa_{\text{e}} \, \kappa_{\text{micro}} (\mu_{\text{e}}+\mu_{\text{micro}})
				+
				8 \mu_{\text{e}} \, \mu_{\text{micro}} (\kappa_{\text{e}} + \kappa_{\text{micro}})
			}
			{
				\mu \, a_2 (\kappa_{\text{e}}+\frac{4}{3} \mu_{\text{e}}) (\kappa_{\text{micro}}+\frac{4}{3} \mu_{\text{micro}})
			}
		},
		\end{align}
	showing that 	  the  positive definiteness of the energy (\ref{eq:energy_RM}) implies that  $f_1$ is a strictly positive real number. Moreover, the function 
	$
	g:(0,\infty)\to \mathbb{R}, \qquad g(x):=
	1-\frac{4 z_{1}^2}{ z_{2} z_{3}}\,\frac{1}{x}
	\tanh \frac{x}{2}
	$
	has the asymptotic behaviour
	\begin{align}
	\lim\limits_{x\to 0}g(x)=1-\frac{2\, z_{1}^2}{ z_{2} z_{3}}=f_1^2>0, \qquad 	\lim\limits_{x\to \infty}g(x)=1
	\end{align}
	and it is monotone increasing since its first derivative is given by
	\begin{align}
	g'(x)=\frac{4 z_{1}^2}{ z_{2} z_{3}}\frac{\sinh x-x}{x^2  (\cosh x+1)}
	\end{align}
	which it is positive for all $x\in(0,\infty)$. Hence, it follows that  due to the positive definiteness of the elastic energy
	\begin{align}
		g(x)>0 \qquad \forall \, x>0,
	\end{align}
	which implies that 
	\begin{align}
		1-\frac{4 z_{1}^2}{f_{1} z_{2} z_{3}}
	\tanh \left(\frac{f_{1} h}{2 L_{\text{c}}}\right)
	\frac{L_{\text{c}}}{h}>0 \qquad \forall \, L_{\text{c}}>0
	\end{align}
	which completes our proof that all the quantities from \eqref{eq:solu_RM} are real and well-defined.

\begin{figure}[H]
	\centering
	\begin{subfigure}{0.49\textwidth}
	    \centering
		\includegraphics[width=\textwidth]{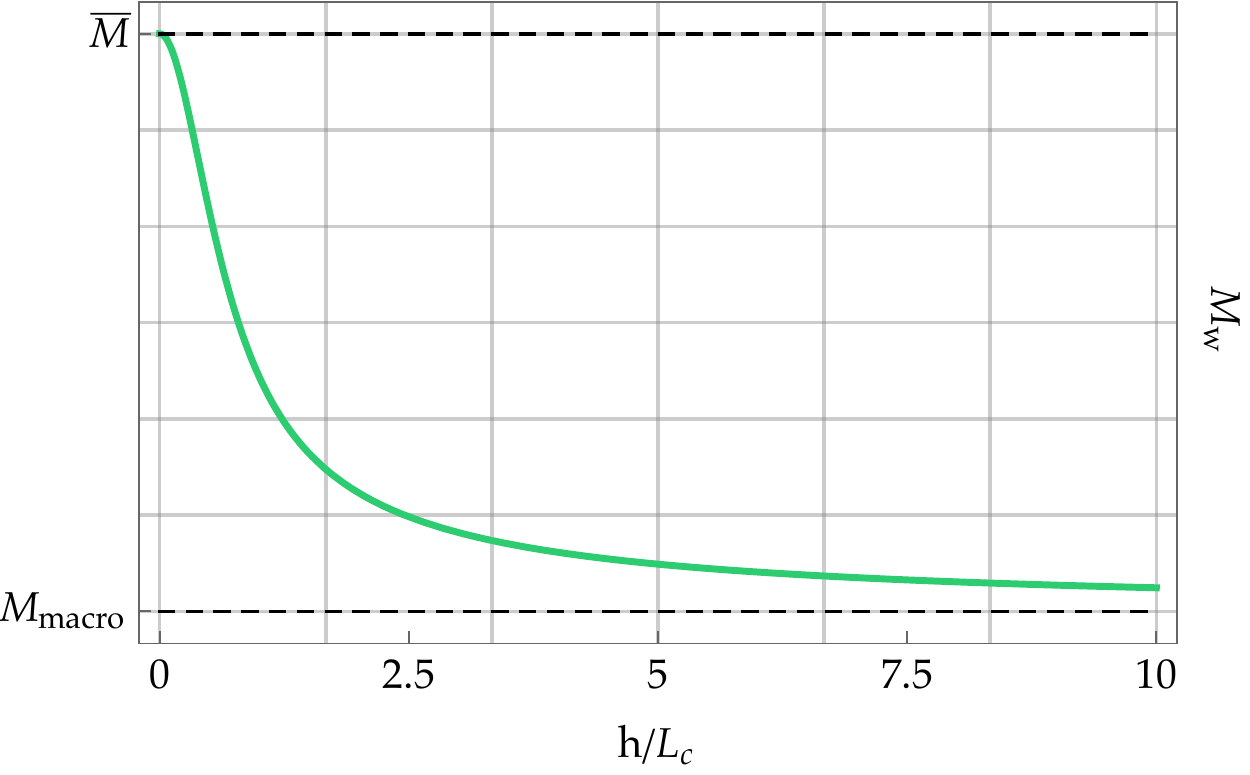}
	\end{subfigure}
	\hfill
	\begin{subfigure}{0.49\textwidth}
		\centering
		\includegraphics[width=0.95\textwidth]{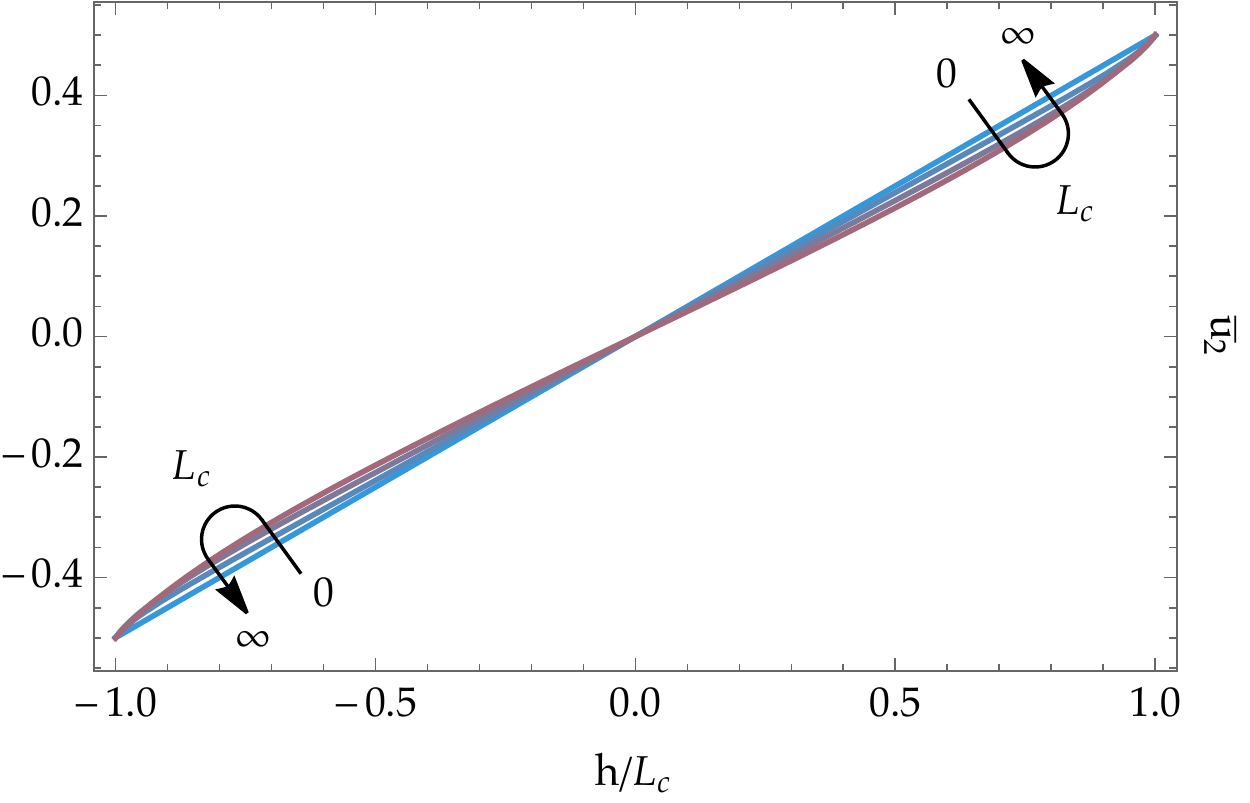}
	\end{subfigure}
	\caption{
	(\textbf{Relaxed micromorphic model}) (\textit{left}) Extensional stiffness $M_{\text{w}}$ while varying $L_{\text{c}}$. The stiffness is bounded as $L_{\text{c}} \to \infty$ ($h\to 0$). The values of the parameters used are: $\mu = 1$, $\lambda_{\text{e}}= 1$, $M_{\text{e}}= 2$, $\lambda_{\text{micro}}= 3$, $M_{\text{micro}}= 4$, $a_2= 1/5$;
	(\textit{right}) Displacement profile across the thickness of the dimensionless $\overline{u}_2=u_2/\left(\gamma \, h\right)$ for different values of $L_{\text{c}}=\{0, 0.014, 0.0\overline{3}, 0.1\}$. The values of the other parameters used in order to maximize the non-homogeneous behaviour are $\mu = 1$, $\lambda_{\text{e}}=1$, $M_{\text{e}}=1$, $\lambda_{\text{micro}}=0.001$, $M_{\text{micro}}= 0.056$, $a_2=0.\overline{3}$.
	}
	\label{fig:stiff_RM}
\end{figure}

The strain energy associated with this solution is
\begin{align}
W(\boldsymbol{\gamma})
=
&
\int_{-h/2}^{h/2} W(\text{D}\boldsymbol{u},\boldsymbol{P},\text{Curl}\, \boldsymbol{P})
\\*
\,
=
&
\frac{1}{2}
\Bigg[
\frac{
\mu \, a_{2}
\left(
\frac{f_{1} z_{1} }{z_{3}}
\right)^2
\left(
\frac{1}{f_{1}}
\sinh \left(\frac{f_{1} h}{L_{\text{c}}}\right)
\frac{L_{\text{c}}}{h}
-1
\right)
}{
\left(
1
-\frac{4 z_{1}^2}{f_{1} z_{2} z_{3}}
\tanh \left(\frac{f_{1} h}{2 L_{\text{c}}}\right)
\frac{L_{\text{c}}}{h}
\right)^2
\cosh^2\left(\frac{f_{1} \, h}{2 L_{\text{c}}}\right)
}
+
\frac{
\cosh ^2\left(\frac{f_{1} h}{2 L_{\text{c}}}\right)
-\frac{z_{1}^2}{z_{2} z_{3}}
\left(
\frac{3}{f_{1}}
\sinh \left(\frac{f_{1} h}{L_{\text{c}}}\right)
\frac{L_{\text{c}}}{h}
-1
\right)
}{
\left(
1
-\frac{4 z_{1}^2}{f_{1} z_{2} z_{3}}
\tanh \left(\frac{f_{1} h}{2 L_{\text{c}}}\right)
\frac{L_{\text{c}}}{h}
\right)^2
\cosh^2\left(\frac{f_{1} \, h}{2 L_{\text{c}}}\right)
}
\notag
\\*
& 
\times
\frac{z_{2}}{z_{3}}
\left(
M_{\text{e}} + M_{\text{micro}} + \lambda_{\text{e}} + \lambda_{\text{micro}} - \frac{2\lambda_{\text{e}}^2}{M_{\text{e}}} - \frac{2\lambda_{\text{micro}}^2}{M_{\text{micro}}}
\right)
\Bigg]
h \, \boldsymbol{\gamma}^2
=
\frac{1}{2} \, M_{\text{w}} \, h \, \boldsymbol{\gamma}^2
\, .
\notag
\end{align}
The plot of the extensional stiffness $M_{\text{w}}$ while varying $L_{\text{c}}$ is shown in Fig.~\ref{fig:stiff_RM}.
The values of $M_{\text{macro}}$ and $M_{\text{micro}}$ are
\begin{align}
    M_{\text{macro}}
    =
    \lim_{L_{\text{c}}\to 0}
    M_{\text{w}}
    &=
    \frac{M_{\text{e}}^2 M_{\text{micro}}+M_{\text{e}} \left(-2 \lambda_{\text{micro}}^2+M_{\text{micro}}^2+M_{\text{micro}} (\lambda_{\text{e}}+\lambda_{\text{micro}})\right)-2 \lambda_{\text{e}}^2 M_{\text{micro}}}{(M_{\text{e}}-\lambda_{\text{e}}-\lambda_{\text{micro}}+M_{\text{micro}}) (M_{\text{e}}+2 (\lambda_{\text{e}}+\lambda_{\text{micro}})+M_{\text{micro}})}
    \notag
    \\*
    &=
    \frac{\kappa_{\text{e}} \, \kappa_{\text{micro}}}{\kappa_{\text{e}}+\kappa_{\text{micro}}}
    +\frac{4}{3}\frac{\mu_{\text{e}} \, \mu_{\text{micro}}}{\mu_{\text{e}}+\mu_{\text{micro}}}
    =
    \kappa_{\text{macro}} +\frac{4}{3} \mu_{\text{macro}}
    =M_{\text{macro}}
    \, ,
    \label{eq:mu_kappa_macro}
    \\*
    \overline{M}
    =
    \lim_{L_{\text{c}}\to\infty}
    M_{\text{w}}
    &=
    \frac{
    M_{\text{e}} \, M_{\text{micro}}
    }{
    M_{\text{e}} + M_{\text{micro}}
    }
    <
    \begin{cases}
    M_{\text{micro}}\\
    M_{\text{e}}
    \end{cases}
    \, ,
    \notag
\end{align}
where $M_{i}=\kappa_{i} + \frac{4}{3}\mu_{i}$ and $\lambda_{i}=\kappa_{i}-\frac{2}{3}\mu_{i}$ with $i=\{\text{macro},\text{micro},\text{e}\}$.
\!\!\!\!\!\!
\footnote{
For the sake of completeness are reported here also the relations between the Young's modulus $E_{i}$ and the Poisson's ratio $\nu_{i}$ in terms of $\kappa_{i}$ and $\mu_{i}$: $E_{i}=\frac{9\kappa_{i}\,\mu_{i}}{3\kappa_{i}+\mu_{i}}$ and $\nu_{i}=\frac{3\kappa_{i}-2\mu_{i}}{2(3\kappa_{i}+\mu_{i})}$ with $i=\{\text{macro},\text{micro},\text{e}\}$.
}

It is highlighted that the structure $\frac{(\bullet)_{\text{e}} \, (\bullet)_{\text{micro}}}{(\bullet)_{\text{e}} + (\bullet)_{\text{micro}}}$ is applicable to evaluate the \textit{macro} coefficients only for the shear and bulk modulus because of the orthogonal energy decomposition ``sym dev/tr'' of which they are related, and especially here it would be a mistake to use this structure for the coefficient $M_{\text{macro}}$ since it will give the value at the \textit{micro-scale}.
For more details about $\lim\limits_{L_{\text{c}}\to\infty} M_{\text{w}}$ see Appendix \ref{app:lim_lc_inf_RMM}.
\subsection{Uniaxial extension problem for the isotropic relaxed micromorphic model with $\nu_e=\nu_{\text{micro}}=0$}
\label{sec:Relaxed_micro_2}
A vanishing Poisson's ratio at the meso- and micro-scale ($\nu_e=\nu_{\text{micro}}=0$) corresponds to a vanishing first Lamé parameter ($\lambda_{\text{e}}=\lambda_{\text{micro}}=0$).
It is easy to see from eq.(\ref{eq:coeff_z_RM}) and eq.(\ref{eq:solu_RM}) that these conditions correspond to 
\begin{align}
    \lambda_{\text{e}}=\lambda_{\text{micro}}=0
    \qquad
    \Longleftrightarrow
    \qquad
    \begin{cases}
    z_1
    =
    0
    \, ,
    \\*[2mm]
    z_2
    =
    \dfrac{
    M_{\text{e}} \, M_{\text{micro}}
    }{
    M_{\text{e}}+M_{\text{micro}}
    }
    =
    \dfrac{
    2\mu_{\text{e}} \, \mu_{\text{micro}}
    }{
    \mu_{\text{e}}+\mu_{\text{micro}}
    }
    \, ,
    \\*[4mm]
    z_3
    =
    M_{\text{e}}+M_{\text{micro}}
    =
    2\left(\mu_{\text{e}}+\mu_{\text{micro}}\right)
    \, ,
    \end{cases}
    \label{eq:coeff_z_RM_l0}
\end{align}
with $M_{i}=\lambda_{i} + 2\mu_{i} = 2\mu_{i}$ with $i=\{\text{micro},\text{e}\}$.
Since the non-linear terms in the solution eq.(\ref{eq:solu_RM}) vanish, we retrieve
\begin{align}
    u_{2} (x_{2})= \boldsymbol{\gamma} \, x_2\, ,
    \qquad\qquad\qquad
    P_{22} (x_{2})= \frac{\mu_{\text{e}}}{\mu_{\text{e}} + \mu_{\text{micro}}} \boldsymbol{\gamma} \, ,
    \qquad\qquad\qquad
    P_{11} (x_{2})=P_{33} (x_{2})= 0\, ,
    \label{eq:solu_RM_l0}
\end{align}
which is a homogeneous elastic solution satisfying the equilibrium equation in the case of a constant micro-distortion tensor $\overline{\boldsymbol{P}}$ (see the Appendix D of \cite{rizzi2021torsion} for further details)
\begin{align}
    \overline{\boldsymbol{P}}
	=
	\frac{\mu_{\text{e}}}{\mu_{\text{e}}+\mu_{\text{micro}}} \, 
	\left(
	\frac{1}{\left | \Omega \right |} \,
	\int_{\Omega}
	\boldsymbol{\text{D}u}
	\, \text{d}V
	\right) \, .
\end{align}
The strain energy associated with this solution is
\begin{equation}
W(\boldsymbol{\gamma})
=
\int_{-h/2}^{h/2} W(\text{D}\boldsymbol{u})
=
\frac{1}{2}
\frac{2\mu_{\text{e}} \, \mu_{\text{micro}}}{\mu_{\text{e}}+\mu_{\text{micro}}} h \, \boldsymbol{\gamma}^2
=
\frac{1}{2} \, M_{\text{macro}} \, h \, \boldsymbol{\gamma}^2
\, ,
\end{equation}
where $M_{\text{macro}} = 2\mu_{\text{macro}} + \lambda_{\text{macro}} = 2\mu_{\text{macro}} = \frac{2\mu_{\text{e}} \, \mu_{\text{micro}}}{\mu_{\text{e}}+\mu_{\text{micro}}}$ is the macro extensional stiffness, since $\lambda_{\text{macro}}=\nu_{\text{macro}}=0$.
\section{Uniaxial extension problem for the isotropic micro-stretch model in dislocation format}

In the micro-stretch model in dislocation format \cite{neff2014unifying,scalia2000extension,de1997torsion,neff2009mean,kirchner2007mechanics}, the micro-distortion tensor $\boldsymbol{P}$ is devoid  from the deviatoric component $\mbox{dev} \, \mbox{sym} \, \boldsymbol{P} = 0 \Leftrightarrow \boldsymbol{P} = \boldsymbol{A} + \omega \boldsymbol{\mathbbm{1}}$, $\boldsymbol{A} \in \mathfrak{so}(3)$, $\omega \in \mathbb{R}$.
The expression of the strain energy for this model in dislocation format can be written as \cite{neff2014unifying}:
\begin{align}
	W \left(\boldsymbol{\mbox{D}u}, \boldsymbol{A},\omega,\mbox{Curl}\,\left(\boldsymbol{A} - \omega \boldsymbol{\mathbbm{1}}\right)\right) 
	\hspace{-2.5cm}
	&
	\notag
	\\*
	=
	&
	\, \mu_{\tiny \mbox{macro}} \left\lVert \mbox{dev} \, \mbox{sym} \, \boldsymbol{\mbox{D}u} \right\rVert^{2}
	+ \frac{\kappa_{\text{e}}}{2} \mbox{tr}^2 \left(\boldsymbol{\mbox{D}u} - \omega \boldsymbol{\mathbbm{1}} \right) 
	+ \mu_{c} \left\lVert \mbox{skew} \left(\boldsymbol{\mbox{D}u} - \boldsymbol{A} \right) \right\rVert^{2}
	+ \frac{9}{2} \, \kappa_{\tiny \mbox{micro}} \, \omega^2
	\label{eq:energy_MST}
	\\*
	&
	+ \frac{\mu \,L_c^2}{2} \,
	\left(
	a_1 \, \left\lVert \mbox{dev sym} \, \mbox{Curl} \, \boldsymbol{A} \right\rVert^2
	+ a_2 \, \left\lVert \mbox{skew} \,  \mbox{Curl} \, \left(\boldsymbol{A} + \omega \boldsymbol{\mathbbm{1}}\right) \right\rVert^2
	+ \frac{a_3}{3} \, \mbox{tr}^2 \left(\mbox{Curl} \, \boldsymbol{A} \right)
	\right) \, ,
	\notag
\end{align}
since $\mbox{Curl} \left(\omega \boldsymbol{\mathbbm{1}}\right) \in \mathfrak{so}(3)$.
The equilibrium equations, in the absence of body forces,   are then
\begin{align}
	\mbox{Div}\overbrace{\left[
		2\mu_{\tiny \mbox{macro}}\,\mbox{dev}\,\mbox{sym} \, \boldsymbol{\mbox{D}u}
		+ \kappa_{\text{e}} \mbox{tr} \left(\boldsymbol{\mbox{D}u} - \omega \boldsymbol{\mathbbm{1}}\right) \boldsymbol{\mathbbm{1}}
		+ 2\mu_{c}\,\mbox{skew} \left(\boldsymbol{\mbox{D}u} - \boldsymbol{A}\right) \right]}^{\mathlarger{\widetilde{\sigma}}\coloneqq}
	&= \boldsymbol{0} \, ,
	\notag
	\\
	2\mu_{c}\,\mbox{skew} \left(\boldsymbol{\mbox{D}u} - \boldsymbol{A}\right)
	\hspace{10.1cm}
	\notag
	\\*
	-\mu \, L_c^2 \, \mbox{skew} \, \mbox{Curl}\,\left(
	a_1 \, \mbox{dev} \, \mbox{sym} \, \mbox{Curl} \, \boldsymbol{A} \, 
	+ a_2 \, \mbox{skew} \, \mbox{Curl} \, \left(\boldsymbol{A} + \, \omega \boldsymbol{\mathbbm{1}}\right) \,
	+ \frac{a_3}{3} \, \mbox{tr} \left(\mbox{Curl} \, \boldsymbol{A} \right)\boldsymbol{\mathbbm{1}} \, 
	\right) &= \boldsymbol{0} \, ,
	\label{eq:equi_MST}
	\\*
	\mbox{tr}
	\bigg[
	2\mu_{\tiny \mbox{macro}}\,\mbox{dev}\,\mbox{sym} \, \boldsymbol{\mbox{D}u}
	\hspace{10.25cm}
	\notag
	\\*
	+ \kappa_{\text{e}} \mbox{tr} \left(\boldsymbol{\mbox{D}u} - \omega \boldsymbol{\mathbbm{1}}\right) \boldsymbol{\mathbbm{1}}
	- \kappa_{\mbox{\tiny micro}} \mbox{tr} \left( \omega \boldsymbol{\mathbbm{1}}\right) \boldsymbol{\mathbbm{1}}
	-\mu \, L_c^2 \,  a_2 \, \mbox{Curl}\,
	\mbox{skew} \, \mbox{Curl} \, \left(\omega \boldsymbol{\mathbbm{1}} + \boldsymbol{A}\right) 
	\bigg]
	&= \boldsymbol{0} \,.
	\notag
\end{align}
According with the reference system shown in Fig.~\ref{fig:intro}, the ansatz for the displacement and micro-distortion fields is
\begin{alignat}{3}
	\boldsymbol{u}(x_2) &=
	\left(
	\begin{array}{c}
		0 \\
		u_2(x_2) \\
		0 
	\end{array}
	\right)
	\, ,
	&
	\qquad\qquad\qquad
	\boldsymbol{A}(x_2) &= 
	\left(
	\begin{array}{ccc}
		0 & 0 & 0 \\
		0 & 0 & 0 \\
		0 & 0 & 0 \\
	\end{array}
	\right)
	\, ,
	\label{eq:ansatz_MST}
	\\*[2mm]
	\text{D}\boldsymbol{u}(x_2) &=
	\left(
	\begin{array}{ccc}
		0 & 0 & 0 \\
		0 & u_{2,2}(x_2) & 0 \\
		0 & 0 & 0  \\
	\end{array}
	\right)
	\, ,
	&\omega\left(x_2\right) \boldsymbol{\mathbbm{1}} &=
	\left(
	\begin{array}{ccc}
		\omega\left(x_2\right) & 0 & 0 \\
		0 & \omega\left(x_2\right) & 0 \\
		0 & 0 & \omega\left(x_2\right) \\
	\end{array}
	\right)
	\, .
	\notag
\end{alignat}
The boundary conditions at the free surface are then
\begin{align}
u_{2}(x_{2} = \pm h/2) = \pm \frac{\boldsymbol{\gamma} \, h}{2} \, , 
\qquad\qquad\qquad
\omega(x_{2} = \pm h/2) = 0
\, .
\label{eq:BC_MST}
\end{align}

Since the ansatz requires $\boldsymbol{A}=0$, \textit{the micro-stretch model coincides with the micro-void model} which will be presented in the Sec.~\ref{sec:Micro-void}.

\section{Uniaxial extension problem for the isotropic Cosserat continuum}
\label{sec:Cos}

The strain energy for the isotropic Cosserat continuum in dislocation tensor format (curvature energy expressed in terms of $ \text{Curl} \boldsymbol{A}$) can be written as \cite{rizzi2021shear,rizzi2021bending,izadi2021torsional,rueger2018strong,cosserat1909theorie,neff2006cosserat,neff2009new,jeong2009numerical,fantuzzi2018some}
\begin{align}
	W \left(\boldsymbol{\text{D}u}, \boldsymbol{A},\text{Curl}\,\boldsymbol{A}\right) = &
	\, \mu_{\text{macro}} \left\lVert \text{sym} \, \boldsymbol{\text{D}u} \right\rVert^{2}
	+ \frac{\lambda_{\text{macro}}}{2} \text{tr}^2 \left(\boldsymbol{\text{D}u} \right) 
	+ \mu_{\text{c}} \left\lVert \text{skew} \left(\boldsymbol{\text{D}u} - \boldsymbol{A} \right) \right\rVert^{2}
	\label{eq:energy_Cos}
	\\*
	&
	+ \frac{\mu \, L_{\text{c}}^2}{2}
	\left(
	a_1 \, \left \lVert \text{dev} \, \text{sym} \, \text{Curl} \, \boldsymbol{A}\right \rVert^2 \, 
	+ a_2 \, \left \lVert \text{skew} \, \text{Curl} \, \boldsymbol{A}\right \rVert^2 \, 
	+ \frac{a_3}{3} \, \text{tr}^2 \left(\text{Curl} \, \boldsymbol{A} \right)
	\right)  \, ,
	\notag
\end{align}
where $\boldsymbol{A} \in \mathfrak{so}(3)$.
The equilibrium equations, in the absence of body forces,   are therefore the following
\begin{align}
	\text{Div}\overbrace{\left[2\mu_{\text{macro}}\,\text{sym} \, \boldsymbol{\text{D}u} + \lambda_{\text{macro}} \text{tr} \left(\boldsymbol{\text{D}u} \right) \boldsymbol{\mathbbm{1}}
		+ 2\mu_{\text{c}}\,\text{skew} \left(\boldsymbol{\text{D}u} - \boldsymbol{A}\right) \right]}^{\mathlarger{\widetilde{\sigma}}\coloneqq}
	&= \boldsymbol{0} \, ,
	\label{eq:equi_Cos}
	\\
	2\mu_{\text{c}}\,\text{skew} \left(\boldsymbol{\text{D}u} - \boldsymbol{A}\right)
	-\mu \, L_{\text{c}}^2 \, \text{skew} \, \text{Curl}\,
	\left(
	a_1 \, \text{dev} \, \text{sym} \, \text{Curl} \, \boldsymbol{A} \, 
	+ \frac{a_3}{3} \, \text{tr} \left(\text{Curl} \, \boldsymbol{A} \right)\boldsymbol{\mathbbm{1}} \, 
	\right)
	&= \boldsymbol{0} \,.
	\notag
\end{align}

According to the reference system shown in Fig.~\ref{fig:intro} and the ansatz (\ref{eq:ansatz_RM}), which has to be particularized as $\boldsymbol{A} = \text{skew} \, \boldsymbol{P} \in \mathfrak{so}(3)$, the ansatz for the displacement field and the micro-rotation for the Cosserat model is
\begin{align}
	\boldsymbol{u}(x_2) =
	\left(
	\begin{array}{c}
		0 \\
		u_2(x_2) \\
		0 
	\end{array}
	\right)
	\, ,
	\qquad\qquad
	\text{D}\boldsymbol{u}(x_2) =
	\left(
	\begin{array}{ccc}
		0 & 0 & 0 \\
		0 & u_{2,2}(x_2) & 0 \\
		0 & 0 & 0  \\
	\end{array}
	\right)
	\, ,
	\qquad\qquad
	\boldsymbol{A}(x_2) = 
	\left(
	\begin{array}{ccc}
		0 & 0 & 0 \\
		0 & 0 & 0 \\
		0 & 0 & 0 \\
	\end{array}
	\right)
	\, .
	\label{eq:ansatz_Cos}
\end{align}
Since $\boldsymbol{A}=\boldsymbol{0}$, the Cosserat model is not able to catch any non-homogeneous response for the uniaxial extension problem and the classical solution (\ref{eq:class_solu}) is retrieved.

The \textit{couple stress model} \cite{neff2015correct,hadjesfandiari2016comparison,neff2016some,koiter1964couple,ghiba2017variant}, which appears by constraining $\boldsymbol{A}=\text{skew} \, \boldsymbol{\text{D}u} \in \mathfrak{so}(3)$ in the Cosserat model, is also not able to catch a non-homogeneous response for the uniaxial extension problem since, due to the ansatz, we would have $\text{skew} \, \boldsymbol{\text{D}u}=\boldsymbol{0}$ as it can be seen in eq.(\ref{eq:ansatz_Cos}).
\section{Uniaxial extension problem for the isotropic micro-void model in dislocation tensor format}
\label{sec:Micro-void}

\begin{figure}[H]
	\centering
	\includegraphics[height=5.5cm]{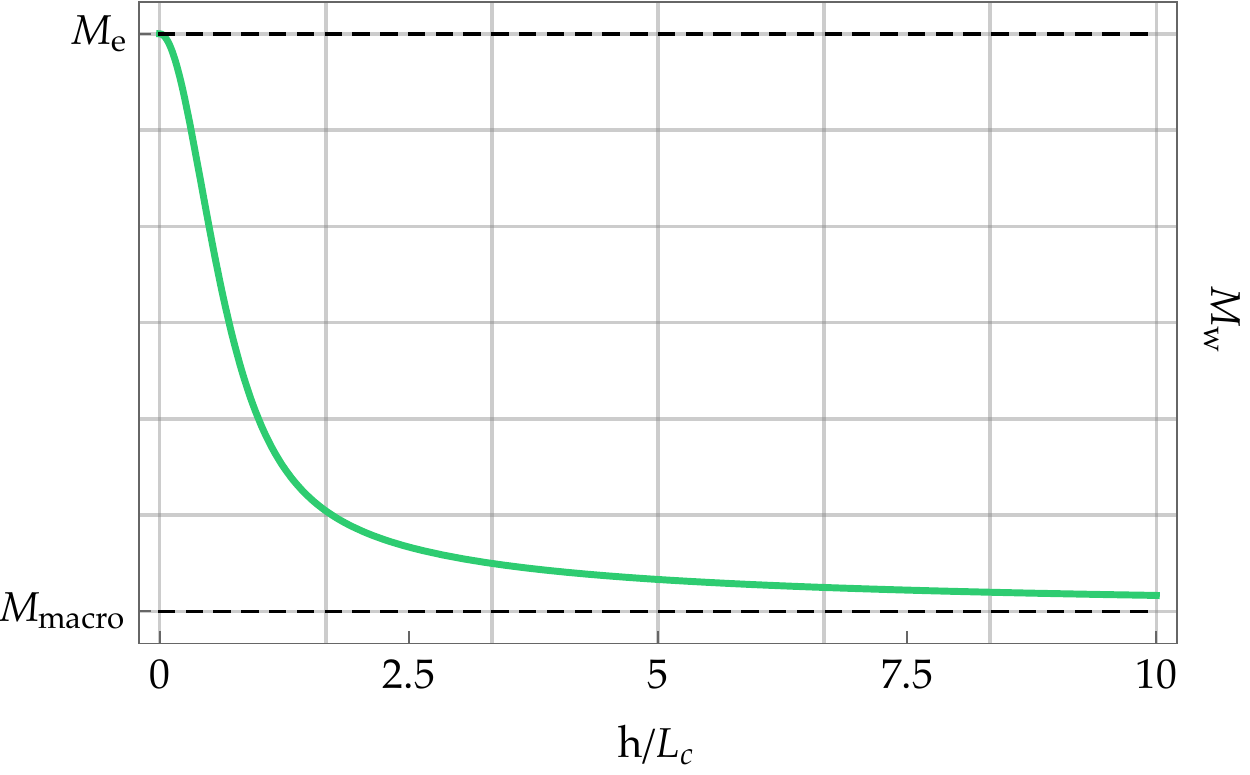}
	\caption{(\textbf{Micro-void model}) Extensional stiffness $M_{\text{w}}$ while varying $L_{\text{c}}$. The stiffness is bounded as $L_{\text{c}} \to \infty$ ($h\to 0$) by $M_{\text{e}}$. The values of the parameters used are: $\mu = 1$, $\lambda_{\text{e}}= 1$, $M_{\text{e}}= 2$, $\kappa_{\text{micro}}= 3$, $a_2= 1/5$.}
	\label{fig:stiff_MV}
\end{figure}

The strain energy for the isotropic micro-void continuum in dislocation tensor format can be obtained from the relaxed micromorphic model by formally letting $\mu_{\text{micro}}\to\infty$ (while keeping $\kappa_{\text{micro}}$ finite) and can be written as \cite{rizzi2021shear,Cowin1983}
\begin{align}
	W \left(\boldsymbol{\text{D}u}, \omega ,\text{Curl}\,\left(\omega \boldsymbol{\mathbbm{1}}\right) \right) = &
	\, \mu_{\text{macro}} \left\lVert \text{dev} \,\text{sym} \, \boldsymbol{\text{D}u}\right\rVert^{2}
	+ \frac{\kappa_{\text{e}}}{2} \text{tr}^2 \left(\boldsymbol{\text{D}u} - \omega \boldsymbol{\mathbbm{1}} \right) 
	+ \frac{\kappa_{\text{micro}}}{2} \text{tr}^2 \left(\omega \boldsymbol{\mathbbm{1}} \right)
	\label{eq:energy_MV}
	\\*
	&
	+ \frac{\mu \,L_{\text{c}}^2 }{2} \,
	a_2 \, \left\lVert \text{Curl} \, \left(\omega \boldsymbol{\mathbbm{1}}\right)\right\rVert^2 
	\, .
	\notag
\end{align}
Here, $\omega : \mathbb{R}^3 \to \mathbb{R}$ is the additional scalar micro-void degree of freedom \cite{Cowin1983}.
The equilibrium equations, in the absence of body forces, are
\begin{align}
	\text{Div}\overbrace{\left[
		2\mu_{\text{macro}} \, \text{dev} \, \text{sym} \, \boldsymbol{\text{D}u}
		+ \kappa_{\text{e}} \text{tr} \left(\boldsymbol{\text{D}u} - \omega \boldsymbol{\mathbbm{1}} \right) \boldsymbol{\mathbbm{1}}
		\right]}^{\mathlarger{\widetilde{\sigma}}\coloneqq}
	&= \boldsymbol{0},
	\label{eq:equi_MV}
	\\
	\frac{1}{3}\text{tr}
	\left[
	\widetilde{\sigma}
	- \kappa_{\text{micro}} \text{tr} \left(\omega \boldsymbol{\mathbbm{1}}\right) \boldsymbol{\mathbbm{1}}
	- \mu \, L_{\text{c}}^{2} \, a_2 \, \text{Curl} \,
	\text{Curl} \, \left(\omega \boldsymbol{\mathbbm{1}}\right)
	\right] &= 0.
	\notag
\end{align}
and the positive definiteness conditions are
\begin{gather}
\label{eq:posi_defe_MV}
    \mu_{\text{macro}} > 0,
    \qquad
    \kappa_{\text{e}} > 0,
    \qquad
    \kappa_{\text{micro}} > 0,
    \qquad
    \mu > 0,
    \qquad
    L_{\text{c}} > 0,
    \qquad a_2 > 0
    \, .
\end{gather}
According with the reference system shown in Fig.~\ref{fig:intro}, the ansatz for the displacement field and the function $\omega(x_2)$ have to be
\begin{align}
	\boldsymbol{u}(x_1,x_2) &=
	\left(
	\begin{array}{c}
		-x_2 \, x_3 \\
		x_1 \, x_3 \\
		0 
	\end{array}
	\right) \, ,
	\qquad
	\omega\left(x_2\right) \boldsymbol{\mathbbm{1}} =
	\left(
	\begin{array}{ccc}
		\omega\left(x_2\right) & 0 & 0 \\
		0 & \omega\left(x_2\right) & 0 \\
		0 & 0 & \omega\left(x_2\right) \\
	\end{array}
	\right) 
	\, ,
	\label{eq:ansatz_MV}
	\\*
	\text{D}\boldsymbol{u}(x_2) &=
	\left(
	\begin{array}{ccc}
		0 & 0 & 0 \\
		0 & u_{2,2}(x_2) & 0 \\
		0 & 0 & 0  \\
	\end{array}
	\right)
	\, .
	\notag
\end{align}
The boundary conditions for the uniaxial extension are
\begin{align}
u_{2}(x_{2} = \pm h/2) = \pm \frac{\boldsymbol{\gamma} \, h}{2} \, , 
\qquad\qquad\qquad
\omega(x_{2} = \pm h/2) = 0
\, .
\label{eq:BC_MV}
\end{align}
After substituting the ansatz (\ref{eq:ansatz_MV}) into the equilibrium equations (\ref{eq:equi_MV}) we obtain the following two differential equations
\begin{align}
    \label{eq:equi_MV_2}
    \frac{1}{3} (3 \kappa_{\text{e}}+4 \mu_{\text{macro}}) \, u_{2}''(x_{2})-\kappa_{\text{e}} \, \omega '(x_{2})
    &
    =
    0
    \, ,
    \\*[2mm]
    \frac{2}{3} a_{2} \, \mu  \, L_{\text{c}}^2 \, \omega ''(x_{2})
    + 3 \kappa_{\text{e}} \, u_{2}'(x_{2})
    - 3 (\kappa_{\text{e}}+\kappa_{\text{micro}}) \, \omega (x_{2})
    &
    =
    0
    \, .
    \notag
\end{align}
After applying the boundary conditions eqs.(\ref{eq:BC_MV}), the solution in terms of $u_2(x_2)$ and $\omega(x_2)$ of the system eqs.(\ref{eq:equi_MV_2}) is
\begin{align}
    u_{2} (x_{2})
    &=
    \frac{
    \frac{x_{2}}{h}
    -\frac{z_{1}}{f_{1}}
    \text{sech}\left(\frac{f_{1} h}{2 L_{\text{c}}}\right) \sinh \left(\frac{f_{1}x_{2}}{L_{\text{c}}}\right)
    \frac{L_{\text{c}}}{h}
    }{
    1
    -\frac{2 z_{1}}{f_{1}}
    \tanh \left(\frac{f_{1} h}{2 L_{\text{c}}}\right)
    \frac{L_{\text{c}}}{h}
    }
    h \, \boldsymbol{\gamma}
    \, ,
    \qquad\qquad
    \omega (x_{2})
    =
    \frac{
    z_{2} \left(1-\text{sech}\left(\frac{f_{1} h}{2 L_{\text{c}}}\right) \cosh \left(\frac{f_{1} x_{2}}{L_{\text{c}}}\right)\right)
    }{
    1
    -\frac{2 z_{1}}{f_{1}}
    \tanh \left(\frac{f_{1} h}{2 L_{\text{c}}}\right)
    \frac{L_{\text{c}}}{h}
    }
    \boldsymbol{\gamma}
    \, ,
    \label{eq:solu_MV}
    \\*
    f_1
    &
    \coloneqq
    \sqrt{
    \frac{4 \mu_{\text{macro}} (\kappa_{\text{e}}+\kappa_{\text{micro}})+3 \kappa_{\text{e}} \kappa_{\text{micro}}}{2 \mu \, a_{2} (3 \kappa_{\text{e}}+4 \mu_{\text{macro}})}
    }
    \, ,
    \quad
    z_1
     \coloneqq
    \frac{3 \kappa_{\text{e}}^2}{(\kappa_{\text{e}}+\kappa_{\text{micro}}) (3 \kappa_{\text{e}}+4
   \mu_{\text{macro}})}
    \, ,
    \quad
    z_2
    \coloneqq
    \frac{\kappa_{\text{e}}}{3 (\kappa_{\text{e}}+\kappa_{\text{micro}})}
    \, .
    \qquad
    \notag
\end{align}
where $f_1>0$, $z_1>0$, and $z_2>0$ are strictly positive in order to match the positive definiteness conditions eq.(\ref{eq:posi_defe_MV}), and the same reasoning applied in the relaxed micromorphic model sections still holds.
The strain energy associated with this solution is
\begin{align}
W(\boldsymbol{\gamma})
=
&
\int_{-h/2}^{h/2} W(\text{D}\boldsymbol{u},\boldsymbol{P},\text{Curl}\, \boldsymbol{P})
\\*
\,
=
&
\frac{1}{2}
\Bigg[
\frac{
\mu \, a_{2} \, f_{1}^2 \, z_{2}^2
\left(
\frac{1}{f_{1}}
\sinh \left(\frac{f_{1} h}{L_{\text{c}}}\right)
\frac{L_{\text{c}}}{h}
-1
\right)
}{
\left(
1-
\frac{2 z_{1}}{f_{1}}
\tanh \left(\frac{f_{1} h}{2 L_{\text{c}}}\right)
\frac{L_{\text{c}}}{h}
\right)^2
\cosh ^2\left(\frac{f_{1} h}{2 L_{\text{c}}}\right)
}
+
\frac{
\left(
1
+z1
+\cosh \left(\frac{f_{1} h}{L_{\text{c}}}\right)
-3 \frac{z_{1}}{f_{1}}
\sinh \left(\frac{f_{1} h}{L_{\text{c}}}\right)
\frac{L_{\text{c}}}{h}
\right)
}{
2\left(
1-
\frac{2 z_{1}}{f_{1}}
\tanh \left(\frac{f_{1} h}{2 L_{\text{c}}}\right)
\frac{L_{\text{c}}}{h}
\right)^2
\cosh ^2\left(\frac{f_{1} h}{2 L_{\text{c}}}\right)
}
\notag
\\*
& 
\times
\left(
\frac{\kappa_{\text{e}} \kappa_{\text{micro}}}{\kappa_{\text{e}}+\kappa_{\text{micro}}}
+
\frac{4 \mu_{\text{macro}}}{3}
\right)
\Bigg]
h \, \boldsymbol{\gamma}^2
=
\frac{1}{2} \, M_{\text{w}} \, h \, \boldsymbol{\gamma}^2
\, .
\notag
\end{align}
The plot of the extensional stiffness $M_{\text{w}}$ while varying $L_{\text{c}}$ is shown in Fig.~\ref{fig:stiff_MV}.
\noindent
The values of the extensional stiffness $M_{\text{w}}$ for $L_{\text{c}}\to 0$ and $L_{\text{c}}\to \infty$ are
\begin{align}
    \lim_{L_{\text{c}}\to 0}
    M_{\text{w}}
    &=
    \frac{\kappa_{\text{e}} \, \kappa_{\text{micro}}}{\kappa_{\text{e}}+\kappa_{\text{micro}}}
    +\frac{4}{3}\mu_{\text{macro}}
    =
    \kappa_{\text{macro}} +\frac{4}{3} \mu_{\text{macro}}
    =
    2\mu_{\text{macro}} + \lambda_{\text{macro}} 
    =M_{\text{macro}}
    \, ,
    \\*
    \lim_{L_{\text{c}}\to\infty}
    M_{\text{w}}
    &=
    \kappa_{\text{e}} +\frac{4}{3} \mu_{\text{macro}}
    =
    \kappa_{\text{e}} +\frac{4}{3} \mu_{\text{e}}
    =
    2\mu_{\text{e}} + \lambda_{\text{e}} 
    =
    M_{\text{e}}
    \, ,
    \notag
\end{align}
where $\mu_{\text{macro}}=\mu_{\text{e}}$ for $\mu_{\text{micro}}\to \infty$, according to eq.(\ref{eq:mu_kappa_macro}).
We note that the extensional stiffness remains bounded as $L_{\text{c}}\to\infty$ ($h\to 0$).
\section{Uniaxial extension problem for the classical isotropic micromorphic continuum without mixed terms}
\label{sec:Micro_morphic}
The expression of the strain energy for the classical isotropic micromorphic continuum \cite{mindlin1964micro,eringen1968mechanics} without mixed terms (like $\langle\text{sym} \boldsymbol{P}, \text{sym}\left(\text{D}\boldsymbol{u} -\boldsymbol{P}\right)\rangle$, etc.) and simplified curvature expression \cite{rizzi2021bending,rizzi2021torsion} can be written as:
\begin{align}
    W \left(\text{D}\boldsymbol{u}, \boldsymbol{P}, \text{D}\boldsymbol{P}\right) 
    = &
    \, \mu_{\text{e}} \left\lVert \text{sym} \left(\text{D}\boldsymbol{u} - \boldsymbol{P} \right) \right\rVert^{2}
    + \dfrac{\lambda_{\text{e}}}{2} \text{tr}^2 \left(\text{D}\boldsymbol{u} - \boldsymbol{P} \right)
    + \mu_{\text{c}} \left\lVert \text{skew} \left(\text{D}\boldsymbol{u} - \boldsymbol{P} \right) \right\rVert^{2}
    \notag
    \\*[3mm]
    &
    + \mu_{\text{micro}} \left\lVert \text{sym}\,\boldsymbol{P} \right\rVert^{2}
    + \dfrac{\lambda_{\text{micro}}}{2} \text{tr}^2 \left(\boldsymbol{P} \right)
    \label{eq:energy_MM}
    \\*[3mm]
	&
	+ \frac{\mu \, L_{\text{c}}^2}{2}
	\Bigg(
	\widetilde{a}_1 \, \left\lVert \text{D}\left(\text{dev} \, \text{sym} \, \boldsymbol{P}\right) \right\rVert^2
	+ \widetilde{a}_2 \, \left\lVert \text{D}\left(\text{skew} \, \boldsymbol{P}\right) \right\rVert^2
	+ \frac{2}{9} \, \widetilde{a}_3 \left\lVert \text{D} \left(\text{tr} \left(\boldsymbol{P}\right)\boldsymbol{\mathbbm{1}} \right) \right\rVert^2 \Big)
	\Bigg)
    \notag
\end{align}
while the equilibrium equations without body forces are the following:
\begin{align}
    \text{Div}\overbrace{\left[2\mu_{\text{e}}\,\text{sym} \left(\text{D}\boldsymbol{u} - \boldsymbol{P} \right) + \lambda_{\text{e}} \text{tr} \left(\text{D}\boldsymbol{u} - \boldsymbol{P} \right) \boldsymbol{\mathbbm{1}}
	+ 2\mu_{\text{c}}\,\text{skew} \left(\text{D}\boldsymbol{u} - \boldsymbol{P} \right)\right]}^{\mathlarger{\widetilde{\boldsymbol{\sigma}}}}
    &
    =
    \boldsymbol{0} \, ,
    \label{eq:equiMic_MM}
    \\*[3mm]
    \widetilde{\sigma}
    - 2 \mu_{\text{micro}}\,\text{sym}\,\boldsymbol{P} - \lambda_{\text{micro}} \text{tr} \left(\boldsymbol{P}\right) \boldsymbol{\mathbbm{1}}
    \hspace{7cm}
    &
    \notag
	\\*
	+\mu L_{\text{c}}^{2} \,
	\text{Div}
	\left[
	\widetilde{a}_1 \, \text{D} \left(\text{dev} \, \text{sym} \, \boldsymbol{P}\right)
	+ \widetilde{a}_2 \, \text{D} \left(\text{skew} \, \boldsymbol{P}\right)
	+ \frac{2}{9} \, \widetilde{a}_3 \, \text{D} \left(\text{tr} \left(\boldsymbol{P}\right)\boldsymbol{\mathbbm{1}} \right)
	\right]
	&
	=
	\boldsymbol{0} \, ,
	\notag
\end{align}
where ($\mu_{\text{e}}$,$\kappa_{\text{e}}=\lambda_{\text{e}}+2/3 \, \mu_{\text{e}}$), ($\mu_{\text{micro}}$,$\kappa_{\text{micro}}=\lambda_{\text{micro}}+2/3 \, \mu_{\text{micro}}$), $\mu_{\text{c}}$, $L_{\text{c}} > 0$, and ($\widetilde{a}_1$,$\widetilde{a}_2$,$\widetilde{a}_3$)$ > 0$ in order to guarantee the positive definiteness of the energy.
According with the reference system shown in Fig.~\ref{fig:intro}, the ansatz for the displacement field and the classical micromorphic model is
\begin{align}
	\boldsymbol{u}(x_2) &=
	\left(
	\begin{array}{c}
		0 \\
		u_{2}(x_{2}) \\
		0 
	\end{array}
	\right)
	\, ,
	\label{eq:ansatz_MM}
	\qquad\qquad
	\boldsymbol{P}(x_2) =
	\left(
	\begin{array}{ccc}
		P_{11}(x_2) & 0 & 0 \\
		0 & P_{22}(x_2) & 0 \\
		0 & 0 & P_{33}(x_2)  \\
	\end{array}
	\right)
	\, ,
	\\*
	\text{D}\boldsymbol{u}(x_2) &=
	\left(
	\begin{array}{ccc}
		0 & 0 & 0 \\
		0 & u_{2,2}(x_2) & 0 \\
		0 & 0 & 0  \\
	\end{array}
	\right)
	\, .
	\notag
\end{align}
The boundary conditions for the uniaxial extension are assumed to be
\begin{gather}
u_{2}(x_{2} = \pm h/2) = \pm \frac{\boldsymbol{\gamma} \, h}{2} \, ,
\qquad\qquad\qquad
\boldsymbol{P}(x_{2} = \pm h/2)  = 0 \, .
\label{eq:BC_MM}
\end{gather}

The calculations are deferred to the micro-strain model Sec.\ref{sec:Micro_strain} since the ansatz, the equilibrium equations, and the boundary conditions are the same, therefore the solution will also be the same.
\section{Uniaxial extension problem for the micro-strain model without mixed terms}
\label{sec:Micro_strain}

The micro-strain model \cite{forest2006nonlinear,hutter2015micromorphic,shaat2018reduced} is the classical Mindlin-Eringen \cite{eringen1968mechanics,mindlin1964micro} model particular case in which it is assumed a priori that the micro-distortion remains symmetric, $\boldsymbol{P}=\boldsymbol{S}\in \text{Sym}(3)$.

The strain energy which we consider is \cite{rizzi2021bending,rizzi2021torsion}
\begin{align}
	W \left(\boldsymbol{\text{D}u}, \boldsymbol{S}, \boldsymbol{\text{D}S}\right) 
	=&
	\, \mu_{\text{e}} \left\lVert \left(\text{sym} \, \boldsymbol{\text{D}u} - \boldsymbol{S}\right) \right\rVert^{2}
	+ \frac{\lambda_{\text{e}}}{2} \text{tr}^2 \left(\boldsymbol{\text{D}u} - \boldsymbol{S} \right)
	+ \mu_{\text{micro}} \left\lVert \, \boldsymbol{S} \right\rVert^{2} 
	+ \frac{\lambda_{\text{micro}}}{2} \text{tr}^2 \left(\boldsymbol{S} \right)
	\label{eq:energy_MS}
	\\*
	&
	+ \frac{\mu \, L_{\text{c}}^2}{2} \,
	\left(
	\widetilde{a}_1 \, \left\lVert \boldsymbol{\text{D}} \left(\text{dev} \, \boldsymbol{S}\right) \right\rVert^2
	+ \frac{2}{9} \, \widetilde{a}_3 \, \left\lVert \text{D} \left(\text{tr} \left(\boldsymbol{S}\right)\boldsymbol{\mathbbm{1}} \right) \right\rVert^2
	\right)
	\, .
	\notag
\end{align}
The chosen 2-parameter curvature expression represents a simplified isotropic curvature (the full isotropic curvature for the micro-strain model would still count 8 parameters \cite{barbagallo2016transparent}).

The equilibrium equations, in the absence of body forces, are therefore the following 
\begin{align}
	\text{Div}\overbrace{\left[
		2\mu_{\text{e}} \, \left(\text{sym} \, \boldsymbol{\text{D}u} - \boldsymbol{S}\right)
		+ \lambda_{\text{e}} \, \text{tr} \left(\boldsymbol{\text{D}u} - \boldsymbol{S} \right) \boldsymbol{\mathbbm{1}}
		\right]}^{\mathlarger{\widetilde{\sigma}}\coloneqq}
	= \boldsymbol{0},
	\notag
	\\*
	2\mu_{\text{e}} \, \left(\text{sym} \, \boldsymbol{\text{D}u} - \boldsymbol{S}\right)
	+ \lambda_{\text{e}} \, \text{tr} \left(\boldsymbol{\text{D}u} - \boldsymbol{S} \right) \boldsymbol{\mathbbm{1}}
	- 2 \mu_{\text{micro}} \, \boldsymbol{S}
	- \lambda_{\text{micro}} \, \text{tr} \left(\boldsymbol{S}\right) \boldsymbol{\mathbbm{1}} \, \, 
	\hspace{2.5cm}
	\label{eq:equi_MS}
	\\*
	+ \, \mu \, L_{\text{c}}^{2}\,
	\text{sym} \, \text{Div} \, 
	\left[
	\widetilde{a}_1 \, \text{D} \left(\text{dev} \, \boldsymbol{S}\right)
	+ \frac{2}{9} \, \widetilde{a}_3 \, \text{D} \left(\text{tr} \left(\boldsymbol{S}\right)\boldsymbol{\mathbbm{1}} \right)
	\right]
	= \boldsymbol{0} \, ,
	\notag
\end{align}
where ($\mu_{\text{e}}$,$\kappa_{\text{e}}=\lambda_{\text{e}}+2/3 \, \mu_{\text{e}}$), ($\mu_{\text{micro}}$,$\kappa_{\text{micro}}=\lambda_{\text{micro}}+2/3 \, \mu_{\text{micro}}$), $L_{\text{c}} > 0$, and ($\widetilde{a}_1$,$\widetilde{a}_3$)$ > 0$ in order to guarantee the positive definiteness of th energy.
The boundary conditions for the uniaxial extension are assumed to be
\begin{gather}
u_{2}(x_{2} = \pm h/2) = \pm \frac{\boldsymbol{\gamma} \, h}{2} \, ,
\qquad\qquad\qquad
\boldsymbol{S}(x_{2} = \pm h/2)  = 0 \, .
\label{eq:BC_MS}
\end{gather}

According with the reference system shown in Fig.~\ref{fig:intro}, the ansatz for the displacement field and the micro-distortion is (which coincides with the classical micromorphic model eq.(\ref{eq:ansatz_MM}))
\begin{align}
	\boldsymbol{u}(x_2) &=
	\left(
	\begin{array}{c}
		0 \\
		u_{2}(x_{2}) \\
		0 
	\end{array}
	\right)
	\, ,
	\label{eq:ansatz_MS}
	\qquad\qquad
	\boldsymbol{S}(x_2) =
	\left(
	\begin{array}{ccc}
		S_{11}(x_2) & 0 & 0 \\
		0 & S_{22}(x_2) & 0 \\
		0 & 0 & S_{33}(x_2)  \\
	\end{array}
	\right)
	\, ,
	\\*
	\text{D}\boldsymbol{u}(x_2) &=
	\left(
	\begin{array}{ccc}
		0 & 0 & 0 \\
		0 & u_{2,2}(x_2) & 0 \\
		0 & 0 & 0  \\
	\end{array}
	\right)
	\, .
	\notag
\end{align}
After substituting the ansatz (\ref{eq:ansatz_MS}) into the equilibrium equations (\ref{eq:equi_MS}) we obtain the following four differential equations
\begin{align}
    M_{\text{e}} \left(u_{2}''(x_{2})-P_{22}'(x_{2})\right)-\lambda_{\text{e}} \left(P_{11}'(x_{2})+P_{33}'(x_{2})\right)
    &
    =
    0
    \, ,
    \notag
    \\*[2mm]
    -\frac{2}{9} \mu \, L_{\text{c}}^2 (3 \widetilde{a}_{1}+\widetilde{a}_{3}) P_{11}''(x_{2})
    +\frac{1}{9} \mu \, L_{\text{c}}^2 (3 \widetilde{a}_{1}-2 \widetilde{a}_{3}) \left(P_{22}''(x_{2})+P_{33}''(x_{2})\right)
    \notag
    \\*
    +(M_{\text{e}}+M_{\text{micro}}) P_{11}(x_{2})
    +(\lambda_{\text{e}}+\lambda_{\text{micro}}) (P_{22}(x_{2})+P_{33}(x_{2}))
    -\lambda_{\text{e}} u_{2}'(x_{2})
    &
    =
    0
    \, ,
    \notag
    \\*[2mm]
    \frac{1}{9} \mu \, L_{\text{c}}^2
    \left(
    (3 \widetilde{a}_{1}-2 \widetilde{a}_{3}) P_{11}''(x_{2})
    -2 (3 \widetilde{a}_{1}+\widetilde{a}_{3}) P_{22}''(x_{2})
    +(3 \widetilde{a}_{1}-2 \widetilde{a}_{3}) P_{33}''(x_{2})
    \right)
    \label{eq:equi_MS_2}
    \\*
    +(M_{\text{e}}+M_{\text{micro}}) P_{22}(x_{2})
    -M_{\text{e}} u_{2}'(x_{2})
    +(\lambda_{\text{e}}+\lambda_{\text{micro}}) (P_{11}(x_{2})+P_{33}(x_{2}))
    &
    =
    0
    \, ,
    \notag
    \\*[2mm]
    \frac{1}{9} \mu \, L_{\text{c}}^2 \left((3 \widetilde{a}_{1}-2 \widetilde{a}_{3}) 
    \left(P_{11}''(x_{2})+P_{22}''(x_{2})\right)
    -2 (3 \widetilde{a}_{1}+\widetilde{a}_{3}) P_{33}''(x_{2})\right)
    \notag
    \\*
    +(M_{\text{e}}+M_{\text{micro}}) P_{33}(x_{2})
    +(\lambda_{\text{e}}+\lambda_{\text{micro}}) (P_{11}(x_{2})+P_{22}(x_{2}))
    -\lambda_{\text{e}} u_{2}'(x_{2})
    &
    =
    0
    \, .
    \notag
\end{align}
Being careful of substituting the system of differential equation with one in which eq.(\ref{eq:equi_MS_2})$_2$ and eq.(\ref{eq:equi_MS_2})$_4$ are replaced with their sum and their difference, respectively, we have
\begin{align}
    M_{\text{e}} \left(u_{2}''(x_{2})-P_{22}'(x_{2})\right)-\lambda_{\text{e}} f_{p}'(x_{2})
    &
    =
    0
    \, ,
    \notag
    \\*[2mm]
    -\frac{1}{9} \mu \, L_{\text{c}}^2
    \left(
    (3 \widetilde{a}_{1}+4 \widetilde{a}_{3}) f_{p}''(x_{2})
    +2 (2 \widetilde{a}_{3}-3 \widetilde{a}_{1}) P_{22}''(x_{2})
    \right)
    &
    \notag
    \\*
    +f_{p}(x_{2}) (M_{\text{e}}+\lambda_{\text{e}}+\lambda_{\text{micro}}+M_{\text{micro}})
    +2 (\lambda_{\text{e}}+\lambda_{\text{micro}}) P_{22}(x_{2})
    -2 \lambda_{\text{e}} u_{2}'(x_{2})
    &
    =
    0
    \, ,
    \notag
    \\*[2mm]
    \frac{1}{9} \mu \, L_{\text{c}}^2
    \left(
    (3 \widetilde{a}_{1}-2 \widetilde{a}_{3}) f_{p}''(x_{2})
    -2 (3 \widetilde{a}_{1}+\widetilde{a}_{3}) P_{22}''(x_{2})
    \right)
    &
    \label{eq:equi_MS_3}
    \\*
    +(M_{\text{e}}+M_{\text{micro}}) P_{22}(x_{2})
    -M_{\text{e}} u_{2}'(x_{2})
    +f_{p}(x_{2}) (\lambda_{\text{e}}+\lambda_{\text{micro}})
    &
    =
    0
    \, ,
    \notag
    \\*[2mm]
    f_{m}(x_{2}) (M_{\text{e}}-\lambda_{\text{e}}-\lambda_{\text{micro}}+M_{\text{micro}})
    -\widetilde{a}_{1} \, L_{\text{c}}^2 \, f_{m}''(x_{2})
    &
    =
    0
    \, ,
    \notag
\end{align}
where $f_{p} (x_2)\coloneqq P_{11}(x_2)+P_{33}(x_2)$ and $f_{m} (x_2)\coloneqq P_{11}(x_2)-P_{33}(x_2)$.
It is highlighted that eq.(\ref{eq:equi_MS_3})$_4$ is a homogeneous second order differential equation depending only on $f_{m}(x_2)$ with homogeneous boundary conditions eq.(\ref{eq:BC_MS}).

Also here, the fact that eq.(\ref{eq:equi_MS_3})$_4$ is an independent equation has its meaning in the symmetry constraint of the uniaxial extensional problem in the direction along the $x_2$- and $x_3$-axis, which requires that ${P_{11}(x_2)=P_{33}(x_2)}$.

\begin{figure}[H]
	\centering
	\begin{subfigure}{0.49\textwidth}
	    \centering
		\includegraphics[width=\textwidth]{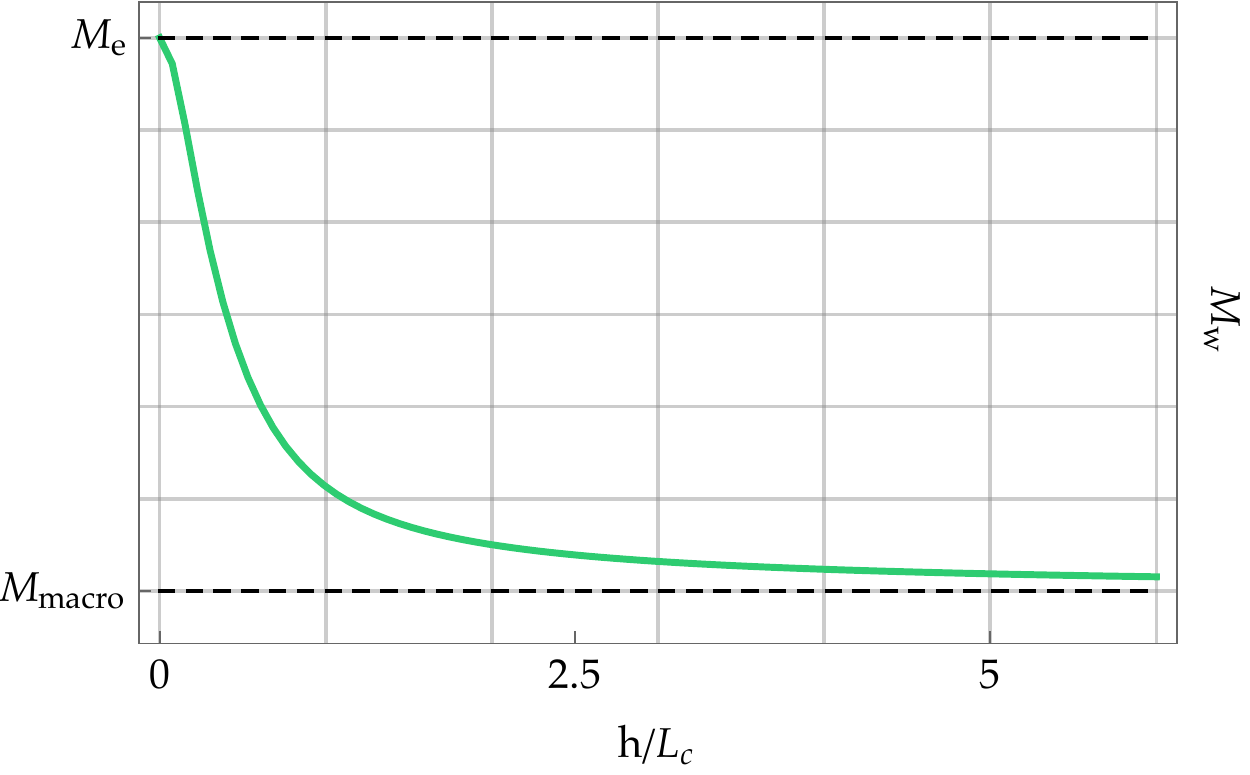}
	\end{subfigure}
	\hfill
	\begin{subfigure}{0.49\textwidth}
		\centering
		\includegraphics[width=0.95\textwidth]{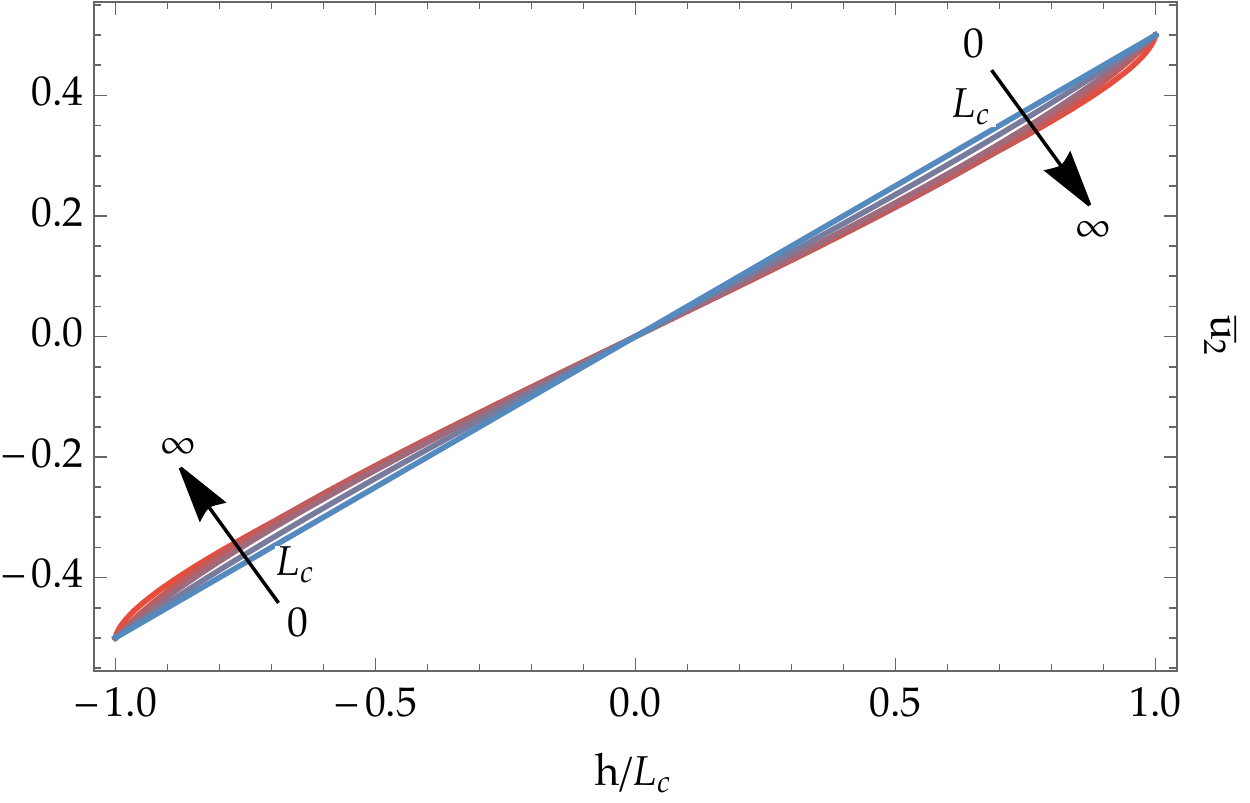}
	\end{subfigure}
	\caption{
	(\textbf{Micro-strain model}) (\textit{left}) Extensional stiffness $M_{\text{w}}$ while varying $L_{\text{c}}$. The stiffness is bounded as $L_{\text{c}} \to \infty$ ($h\to 0$) and converges to $M_{\text{e}}$. The values of the parameters used are: $\mu = 1$, $\lambda_{\text{macro}}= 1$, $M_{\text{macro}}= 3$, $\lambda_{\text{micro}}= 9.69$, $M_{\text{micro}}= 12$, $\widetilde{a}_1= 1/5$, $\widetilde{a}_3= 1/6$;
	(\textit{right}) Displacement profile across the thickness of the dimensionless $\overline{u}_2=u_2/\left(\gamma \, h\right)$ for different values of $L_{\text{c}}=\{0, 3, 5, 10, \infty\}$. The values of the other parameters used in order to maximize the non-homogeneous behaviour are $\mu = 1$, $\lambda_{\text{e}}= 11$, $M_{\text{e}}= 33$, $\lambda_{\text{micro}}= 1.1$, $M_{\text{micro}}= 3.3$, $\widetilde{a}_1= 1$, $\widetilde{a}_3= 1/6$.
	}
	\label{fig:stiff_MM}
\end{figure}

The solution and the measure of the apparent stiffness are too complicated to be reported here, but nevertheless, it is possible to plot how the apparent stiffness behaves while changing $L_{\text{c}}$ (see Fig.~\ref{fig:stiff_MM}).
We note that the extensional stiffness remains bounded as $L_{\text{c}}\to\infty$ ($h\to 0$) and converges to $M_{\text{e}}$.
The solution obtained for the micro-strain model for the uniaxial extension problem also holds for the classical micromorphic problem presented in Sec.\ref{sec:Micro_morphic}.
\section{Uniaxial extension problem for the second gradient continuum}

The strain energy density for the isotropic second gradient with simplified curvature \cite{mindlin1964micro,dell2009generalized,rizzi2021bending,rizzi2021torsion,altenbach2019higher} is
\begin{align}
	W \left(\boldsymbol{\text{D}u}, \boldsymbol{\text{D}^2 u}\right)
	= &
	\, \mu_{\text{macro}} \left\lVert \text{sym}\,\boldsymbol{\text{D}u} \right\rVert^{2}
	+ \frac{\lambda_{\text{macro}}}{2} \text{tr}^2 \left(\boldsymbol{\text{D}u} \right)
	\label{eq:energy_Strain_Grad}
	\\*
	&
	+ \frac{\mu \, L_{\text{c}}^2}{2}
	\left(
	\widetilde{a}_1 \, \left\lVert \text{D} \Big(\text{dev} \, \text{sym} \, \boldsymbol{\text{D} u}\Big) \right\rVert^2
	+ \widetilde{a}_2 \, \left\lVert \text{D} \Big(\text{skew} \, \boldsymbol{\text{D} u}\Big) \right\rVert^2
	+  \frac{2}{9} \, \widetilde{a}_3 \, \left\lVert \text{D}
	\Big( 
	\text{tr} \left(\boldsymbol{\text{D} u}\right) \, \boldsymbol{\mathbbm{1}}
	\Big) \right\rVert^2
	\right)
	\, ,
	\notag
\end{align}
while the equilibrium equations without body forces are the following:
\begin{align}
	\text{Div}\bigg[
	2 \mu_{\text{macro}} \,\text{sym}\,\boldsymbol{\text{D}u}
	+ \lambda_{\text{macro}} \text{tr} \left(\boldsymbol{\text{D}u}\right) \boldsymbol{\mathbbm{1}}
	\hspace{8cm}
	\label{eq:equi_Strain_Grad}
	\\*
	- \mu L_{\text{c}}^{2} \,
	\left(
	\widetilde{a}_1 \, \text{dev} \, \text{sym} \, \boldsymbol{\Delta} \left(\boldsymbol{\text{D}u}\right)
	+ \widetilde{a}_2 \, \text{skew} \, \boldsymbol{\Delta} \left(\boldsymbol{\text{D}u}\right)
	+ \frac{2}{9} \, \widetilde{a}_3 \, \text{tr} \left(\boldsymbol{\Delta} \left(\boldsymbol{\text{D}u}\right)\right)\boldsymbol{\mathbbm{1}}
	\right)
	\bigg]
	= \boldsymbol{0} \, ,
	\notag
\end{align}
where $(\mu_{\text{macro}},\kappa_{\text{macro}},\mu,\widetilde{a}_1,\widetilde{a}_3)>0$ in order to guarantee the positive definiteness of the energy.
Due to the uniaxial extension problem symmetry the following structure of $\boldsymbol{u} = \left(0,u_{2}(x_{2}), 0\right)^T$ has been chosen, which results in having only the component $u_{2,2}$ different from zero in the gradient of the displacement $\text{D}\boldsymbol{u}$.
The boundary conditions for the uniaxial extension are (see Fig.~\ref{fig:intro}) assumed to be
\begin{equation}
u_{2}(x_{2} = \pm h/2) = \pm \frac{\boldsymbol{\gamma} \, h}{2}
\, ,
\qquad\qquad\qquad
u'_{2}(x_{2} = \pm h/2) = 0
\, .
\label{eq:BC_SGM}
\end{equation}
After substituting the expression of the displacement field in eq.(\ref{eq:equi_Strain_Grad}), the non-trivial equilibrium equation reduces to
\begin{equation}
(\lambda_{\text{micro}}+2 \mu_{\text{micro}}) \, u_{2}''(x_{2})-\frac{1}{3} \widetilde{a}_{3} \, \mu \, L_{\text{c}}^2 \,  u_{2}^{(4)}(x_{2}) = 0 \, .
\label{eq:equiShe_Other_SGM}
\end{equation}

\begin{figure}[H]
	\centering
	\begin{subfigure}{0.49\textwidth}
	    \centering
		\includegraphics[width=\textwidth]{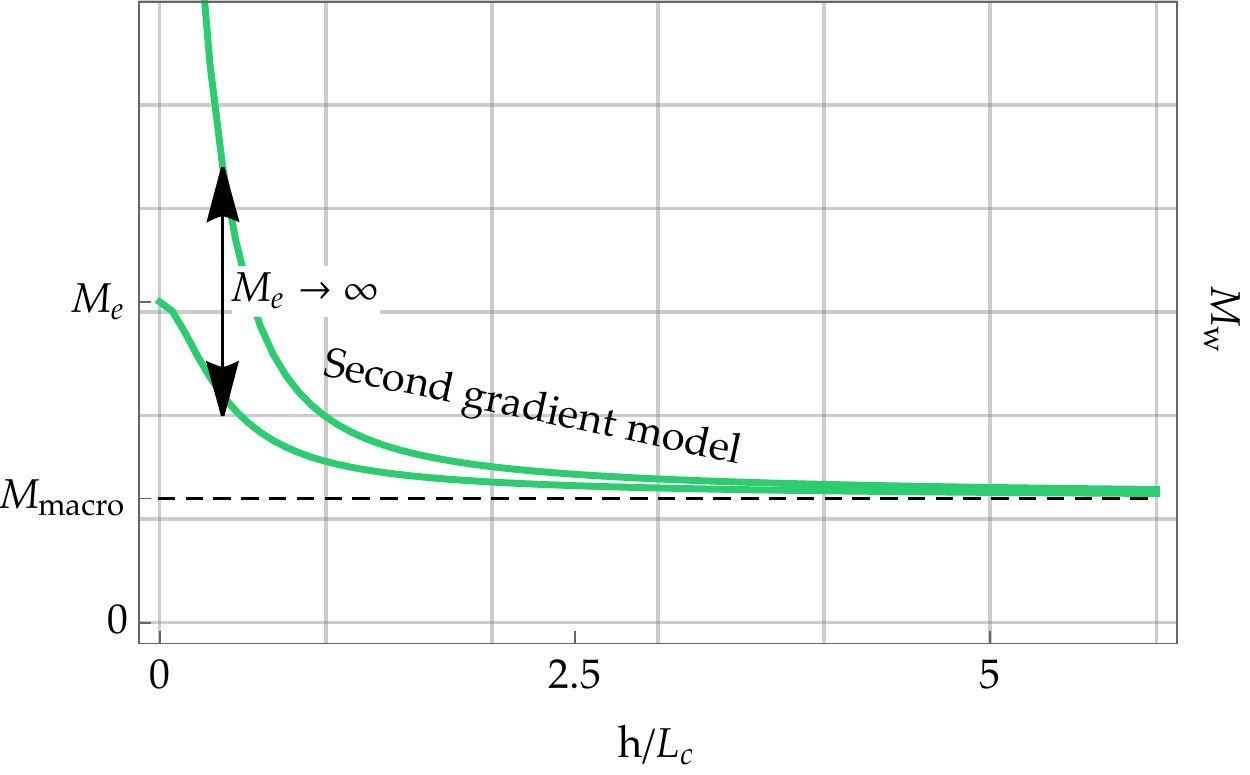}
	\end{subfigure}
	\hfill
	\begin{subfigure}{0.49\textwidth}
		\centering
		\includegraphics[width=0.95\textwidth]{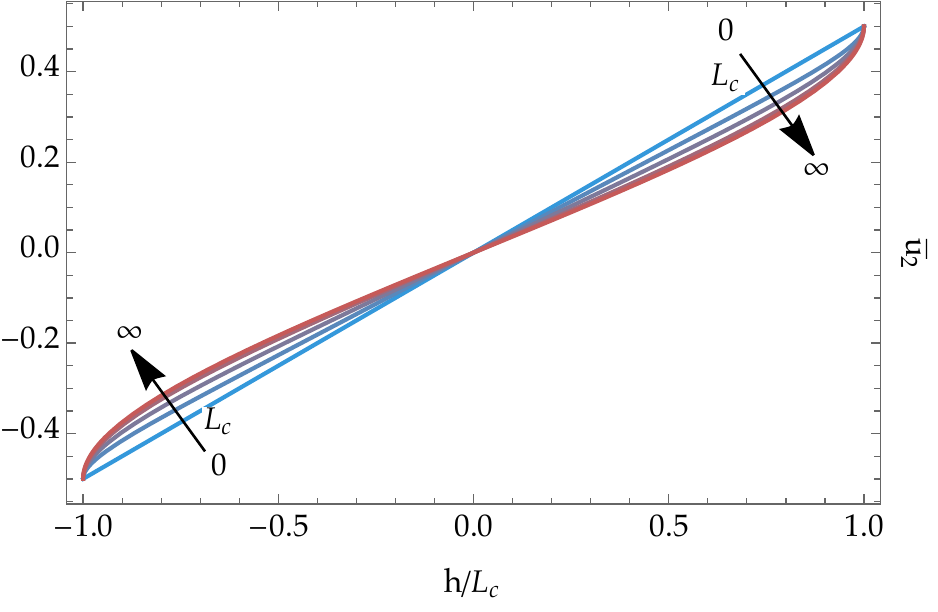}
	\end{subfigure}
	\caption{
	(\textbf{Second gradient model}) (\textit{left}) Extensional stiffness $M_{\text{w}}$ while varying $L_{\text{c}}$. The stiffness is unbounded as $L_{\text{c}} \to \infty$ ($h\to 0$). The values of the parameters used are: $\mu = 1$, $\mu_{\text{macro}}= 1$, $\lambda_{\text{macro}}= 2$, $\widetilde{a}_3= 4$;
	(\textit{right}) Displacement profile across the thickness of the dimensionless $\overline{u}_2=u_2/\left(\gamma \, h\right)$ for different values of $L_{\text{c}}=\{0, 0.1, 0.2, 0.35, \infty\}$. The values of the other parameters used in order to maximize the non-homogeneous behaviour are $\mu = 1$, $\lambda_{\text{micro}}=1$, $M_{\text{micro}}= 1$, $\widetilde{a}_3=2$.
	}
	\label{fig:stiff_SG}
\end{figure}

After applying the boundary conditions to the solution of eq.(\ref{eq:equiShe_Other_SGM}), it results that $u_{2}(x_2)$ is given by \cite{rizzi2019identificationI,rizzi2019identificationII}
\begin{equation}
u_{2}(x_{2}) = 
\frac{
\frac{2x_2}{h}
-
\frac{2}{f_1}
\sinh \left(f_1\frac{x_2}{L_{\text{c}}}\right)
\sech \left(\frac{f_1}{2}\frac{h}{L_{\text{c}}}\right)
\frac{L_{\text{c}}}{h}
}{
1
-
\frac{2}{f_1}
\tanh \left(\frac{f_1}{2}\frac{h}{L_{\text{c}}}\right)
\frac{L_{\text{c}}}{h}
}
\frac{\boldsymbol{\gamma}  h}{2}
\, ,
\qquad\qquad
f_1 := \sqrt{\frac{\lambda_{\text{macro}}+2 \mu_{\text{macro}}}{\mu \, \widetilde{a}_{3}/3 }} \, .
\label{eq:uprime_usecond_SGM}
\end{equation}
where $f_1>0$ is strictly positive in order to match the positive definiteness conditions and the same reasoning applied in the relaxed micromorphic model sections still holds.
The strain energy (\ref{eq:equi_Strain_Grad}) becomes then
\begin{align}
	W(\boldsymbol{\gamma})
    =
    \int_{0}^{h}
    W \left(\boldsymbol{\text{D}u}, \boldsymbol{\text{D}^2 u}\right)
	= 
	\frac{1}{2}
	\left[
	\frac{
	\overbrace{\lambda_{\text{macro}} + 2\mu_{\text{macro}}}^{M_{\text{macro}}}
	}{
	1-\frac{2}{f_1} \tanh \left(\frac{f_1}{2}\frac{h}{L_{\text{c}}}\right)
	\frac{L_{\text{c}}}{h}
	}
	\right]
	h \, \boldsymbol{\gamma}^2
	=
	\frac{1}{2} \, M_{\text{w}} \, h \, \boldsymbol{\gamma}^2
	\, .
	\notag
\end{align}
The plot of the extensional stiffness $M_{\text{w}}$ while varying $L_{\text{c}}$ is shown in Fig.~\ref{fig:stiff_SG}.
\section{Conclusions}
Only the second gradient formulation produces an unbounded apparent stiffness as $L_{\text{c}}\to\infty$ ($h\to0$).
Otherwise, different bounded limit stiffnesses are observed.
The relaxed micromorphic model determines $\overline{M} = \frac{M_{\text{e}} \, M_{\text{micro}}}{M_{\text{e}} + M_{\text{micro}}}$, which is less than $M_{\text{micro}}$ and $M_{\text{e}}$, while the micro-strain model determines $M_{\text{e}}$ as limit stiffness.
The Cosserat model is not able to catch a non-homogeneous solution and provides no size-effect.
The different limit stiffnesses for the relaxed micromorphic model versus the full micromorphic and micro-strain model approach respectively, suggest that the meaning of classical experimental tests does not have an unambiguous deformation and micro-deformation solution field anymore, and this is due to the fact that we can have different boundary conditions on the components of the micro-distortion tensor depending on what each model requires to constrain. This allows the existence of different uniaxial extension-like problems and not just one like for a classical Cauchy material.

{\scriptsize
	\paragraph{{\scriptsize Acknowledgements.}}
	Angela Madeo acknowledges support from the European Commission through the funding of the ERC Consolidator Grant META-LEGO, N° 101001759.
	Angela Madeo and Gianluca Rizzi acknowledge funding from the French Research Agency ANR, “METASMART” (ANR-17CE08-0006).
	Angela Madeo and Gianluca Rizzi acknowledge support from IDEXLYON in the framework of the “Programme Investissement d’Avenir” ANR-16-IDEX-0005.
	Hassam Khan acknowledges the  support of the German Academic Exchange Service (DAAD) and the Higher Education Commission of Pakistan (HEC).
	Ionel Dumitrel Ghiba acknowledges support from a grant of the Romanian Ministry of Research and Innovation, CNCS-UEFISCDI, project number PN-III-PN-III-P1-1.1-TE-2021-0783, within PNCDI III.
	Patrizio Neff acknowledges support in the framework of the DFG-Priority Programme 2256 ``Variational Methods for Predicting Complex Phenomena in Engineering Structures and Materials", Neff 902/10-1, Project-No. 440935806.
}


\begingroup
\setstretch{0.8}
\setlength\bibitemsep{3pt}
\printbibliography
\endgroup


\begin{footnotesize}
\appendix
\section{The Limit $L_{\text{c}} \to \infty$ for the relaxed micromorphic model}
\label{app:lim_lc_inf_RMM}
The limit of the energy, eq.(\ref{eq:energy_RM}), for $L_{\text{c}} \to \infty$,  requires that $ \left\lVert \text{Curl}\,\boldsymbol{P} \right\rVert = 0$, which implies that $\boldsymbol{P} = \text{D} \boldsymbol{\zeta}$, for some $\zeta : \Omega \to \mathbb{R}^3$. The energy eq.(\ref{eq:energy_RM}) now becomes

\begin{align}
W \left(\boldsymbol{\text{D} u}, \boldsymbol{\text{D} \zeta}\right) = &
\, \mu_{\text{e}} \left\lVert \text{sym} \left(\boldsymbol{\text{D} u} - \boldsymbol{\text{D} \zeta} \right) \right\rVert^{2}
+\dfrac{\lambda_{\text{e}}}{2} \text{tr}^2 \left(\boldsymbol{\text{D} u} - \boldsymbol{P} \right) 
+ \mu_{\text{micro}} \left\lVert \text{sym}\,\boldsymbol{\text{D} \zeta} \right\rVert^{2}
+ \dfrac{\lambda_{\text{micro}}}{2} \text{tr}^2 \left(\boldsymbol{\text{D} \zeta} \right) \, ,
\label{eq:energyLimit}
\end{align}
and that eq.(\ref{eq:equi_RM}) turns into
\begin{equation}
\begin{array}{rr}
\text{Div}
\overbrace{
\left[
2\mu_{\text{e}}\,\text{sym} \left(\boldsymbol{\text{D} u} - \boldsymbol{\text{D} \zeta} \right) 
+ \lambda_{\text{e}} \text{tr} \left(\boldsymbol{\text{D} u} - \boldsymbol{\text{D} \zeta} \right) \boldsymbol{\mathbbm{1}}
\right]
}^{
\mathlarger{\widetilde{\boldsymbol{\sigma}}\coloneqq}
}
&= \boldsymbol{0} \, ,
\\*
\widetilde{\sigma}
- 2 \mu_{\text{micro}}\,\text{sym}\,\boldsymbol{\text{D} \zeta} - \lambda_{\text{micro}} \text{tr} \left(\boldsymbol{\text{D} \zeta}\right) \boldsymbol{\mathbbm{1}}
&= \boldsymbol{0} \, ,
\end{array}
\label{eq:equiMicLim}
\end{equation}
with consistent coupling boundary condition $\text{D}\,\boldsymbol{u} \cdot \boldsymbol{\tau} = \text{D} \boldsymbol{\zeta} \cdot \boldsymbol{\tau}$.
Given eq.~(\ref{eq:equiMicLim})$_1$, eq.~(\ref{eq:equiMicLim})$_2$ reduces to be
\begin{equation}
\text{Div}\left[ 
2 \mu_{\text{micro}}\,\text{sym}\,\boldsymbol{\text{D} \zeta} + \lambda_{\text{micro}} \text{tr} \left(\boldsymbol{\text{D} \zeta}\right) \boldsymbol{\mathbbm{1}}
\right] = \boldsymbol{0} \, ,
\label{eq:equiMicLim2}
\end{equation}
which, for the uniaxial extension problem with boundary condition $u_{2} \left(x_2 = \pm h/2 \right) = \pm \boldsymbol{\gamma} \, h/2$, is equivalent to
\begin{equation}
	\text{D} \boldsymbol{\zeta} = 
	\left(\begin{array}{ccc}
	0 & 0 & 0 \\
	0 & a & 0 \\
	0 & 0 & 0 \\
	\end{array}\right) \, ,
	\quad
	\text{D} \boldsymbol{u} = 
	\left(\begin{array}{ccc}
	0 & \boldsymbol{\gamma} & 0 \\
	0 & 0 & 0 \\
	0 & 0 & 0 \\
	\end{array}\right) \, ,
	\label{eq:zeta_u_Lim}
\end{equation}
where $a$ is an arbitrary constant. This solution to eqs.(\ref{eq:equiMicLim}) is therefore \textbf{not} unique.
Inserting $\text{D} \boldsymbol{u}$ and $\text{D} \boldsymbol{\zeta}$ from eq.(\ref{eq:zeta_u_Lim}) in eq.(\ref{eq:energyLimit}), the following energy expression is recovered
\begin{equation}
I \left(a\right) = 
\frac{1}{2}
\left(
2 a^2 M_{\text{micro}}
+2 M_{\text{e}} (a-\boldsymbol{\gamma} )^2
\right)
\, ,
\label{eq:energySubLim}
\end{equation}
which has to be minimized with respect to $a$ in order to remove the non-uniqueness of the equilibrium system eqs.(\ref{eq:equiMicLim}), which means that the following relation
\begin{equation}
\dfrac{\partial}{\partial a}
\left(
a^2 M_{\text{micro}}
+M_{\text{e}} (a-\boldsymbol{\gamma} )^2
\right)
=
2 a (M_{\text{e}}+M_{\text{micro}})
-2 \boldsymbol{\gamma} \, M_{\text{e}}
= 0
 \label{eq:minimiza}
\end{equation}
has to be satisfied.
The solution of eq.(\ref{eq:minimiza}) is 
$a_{\text{min}} = \dfrac{M_{\text{e}}}{M_{\text{e}} + M_{\text{micro}}} \boldsymbol{\gamma}$.
Finally it is possible to substitute $a_{\text{min}}$ into eq.(\ref{eq:zeta_u_Lim}) obtaining
\begin{equation}
\text{D} \boldsymbol{\zeta} = 
\left(\begin{array}{ccc}
0 & 0 & 0 \\
0 & \dfrac{M_{\text{e}}}{M_{\text{e}} + M_{\text{micro}}} \boldsymbol{\gamma} & 0 \\
0 & 0 & 0 \\
\end{array}\right),
\quad
\text{D} \boldsymbol{u} = 
\left(\begin{array}{ccc}
0 & \boldsymbol{\gamma} & 0 \\
0 & 0 & 0 \\
0 & 0 & 0 \\
\end{array}\right).
\label{eq:uLim2}
\end{equation}
The solution eq.(\ref{eq:uLim2}) satisfy the equilibrium equations, the boundary conditions, and the minimum energy requirement.
The expression of the energy now become
\begin{align}
    W(\boldsymbol{\gamma})
    =
    \int_{-h/2}^{h/2}
    W(\text{D} \boldsymbol{u},\text{D} \boldsymbol{\zeta})
    =
    \frac{1}{2}
    \dfrac{M_{\text{e}} \, M_{\text{micro}}}{M_{\text{e}} + M_{\text{micro}}}
    h \, \boldsymbol{\gamma}^2
    =
    \frac{1}{2}
    \overline{M}
    h \, \boldsymbol{\gamma}^2 \, ,
\end{align}
with $\overline{M}=\dfrac{M_{\text{e}} \, M_{\text{micro}}}{M_{\text{e}} + M_{\text{micro}}}$ the extensional stiffness for the relaxed micromorphic when $L_{\text{c}} \to \infty$.
%
\end{footnotesize}
\end{document}